\pdfoutput=1
\documentclass[preprint,12pt]{elsarticle}
\usepackage{amsmath,amssymb,bm}
\usepackage{graphicx}
\usepackage{booktabs}
\usepackage{xcolor}
\usepackage{tikz}
\usetikzlibrary{arrows.meta,calc,decorations.pathmorphing}
\usepackage[colorlinks=true,allcolors=blue!45!black]{hyperref}
\biboptions{sort&compress}
\journal{Journal of Sound and Vibration}

\newcommand{\D}{\mathrm{d}}
\newcommand{\I}{\mathrm{i}}
\newcommand{\E}{\mathrm{e}}
\newcommand{\Ad}{\mathcal{A}}
\newcommand{\Zt}{Z_t}
\newcommand{\kb}{\bar{k}}
\newcommand{\Sr}{\mathit{Sr}}
\newcommand{\Rey}{\mathit{Re}}
\newcommand{\dPWL}{\Delta\mathrm{PWL}}
\newcommand{\dSPL}{\Delta\mathrm{SPL}}
\newcommand{\Hgr}{\mathcal{H}}
\newcommand{\Wup}{\mathcal{W}}
\newcommand{\Dc}{\mathcal{D}}

\makeatletter
\def\Xint#1{\mathchoice
  {\XXint\displaystyle\textstyle{#1}}%
  {\XXint\textstyle\scriptstyle{#1}}%
  {\XXint\scriptstyle\scriptscriptstyle{#1}}%
  {\XXint\scriptscriptstyle\scriptscriptstyle{#1}}%
  \!\int}
\def\XXint#1#2#3{{\setbox0=\hbox{$#1{#2#3}{\int}$}
    \vcenter{\hbox{$#2#3$}}\kern-.5\wd0}}
\makeatother
\newcommand{\pvint}{\Xint-}

\begin{document}

\begin{frontmatter}

\title{Leading-edge noise reduction by perforated and porous inserts:
a compressible finite-chord prediction without fitted constants}

\author{Seongkyu Lee}
\ead{skulee@ucdavis.edu}
\address{Department of Mechanical and Aerospace Engineering, University of
California, Davis, CA 95616, USA}

\begin{abstract}
Porous and perforated leading edges reduce turbulence-interaction noise, but
predictions of the reduction have relied on material impedances adjusted to the
acoustic data. This paper removes that adjustment. The finite-chord scattering
problem is solved in compressible subsonic flow for an arbitrary chordwise
admittance, with the shed wake, the Kutta condition and
permeability-dependent edge singularities, and closed to the far field with the
finite span and the array aperture represented. The admittance of a perforated
insert is derived from the Rayleigh conductivity of one aperture with three
computed corrections: plate thickness, interaction between neighbouring
apertures, and Howe's grazing-flow blockage evaluated at the aperture-averaged
velocity of the boundary layer. A bulk porous insert is represented by
its measured permeability and pore-fluid inertia. For three perforates at four
flow conditions the prediction agrees with measurements to 2.5 dB rms
over 84 comparisons, against 4.4 dB with the unscaled impedance used
previously. Although the Mach number is 0.05, the chord is acoustically
non-compact, and an incompressible solution rises where the measurement falls.
Two of three bulk porous inserts are predicted to within 1.7 dB rms where
interaction noise dominates. Narrow, equally spaced peaks measured on
perforated plates are reproduced by none of the admittances tested; their
spacing implies a source convecting at 0.62 of the free-stream speed, absent
from a frozen-gust model. A parameter study shows that the perforate geometry
acts almost entirely through one low-frequency admittance group, in which the
hole radius cancels for holes buried in the laminar boundary layer.
\end{abstract}


\end{frontmatter}

\section{Introduction}
\label{sec:intro}

Turbulence--aerofoil interaction noise, or leading-edge noise, is generated
when vorticity carried by an oncoming stream is scattered into sound at the
leading edge of a blade or vane. It dominates the broadband spectrum of fans
and open rotors ingesting wakes, of wind turbines operating in atmospheric
turbulence, and of rotors ingesting their own or a neighbouring wake
\citep{moreau2019,glegg2017}. Its prediction rests on the unsteady thin-aerofoil
theory of \citet{sears1941} and on its compressible extension by
\citet{amiet1975}, who converted the gust response of a flat plate into a
far-field spectrum. \citet{ffowcs1970} showed why the leading edge matters so
much: an edge scatters near-field hydrodynamic pressure into sound with an
efficiency controlled by the inverse-square-root singularity of the loading
there, a mechanism since examined by separating the hydrodynamic and acoustic
parts of the pressure near an edge \citep{kang2025prf}. Weakening that singularity is therefore a direct route to noise
reduction, and making the edge permeable is one way to do it.

Experiments have established that permeable leading edges work.
\citet{sarradj2007} and \citet{geyer2010,geyer2011} measured aerofoils made
entirely of porous materials spanning three decades of flow resistivity, and
\citet{geyer2014} examined partially porous surfaces. \citet{roger2013}
compared porous and serrated edges directly. \citet{chaitanya2020jsv} measured
perforated flat plates, both fully perforated and with perforated
leading-edge inserts, and identified an interference between the leading edge
and the end of the insert. \citet{palleja2022} moved the perforation downstream
of a rigid leading edge, a placement intended to limit the aerodynamic penalty,
and found interference peaks set by the distance from the leading edge to the
end of the porous section. \citet{bowen2022} measured additively manufactured
bulk porous leading edges on a NACA~0012 aerofoil with independently measured
permeability, and \citet{hales2024} measured perforated leading-edge inserts of
three open-area ratios in anisotropic cylinder-wake turbulence at four flow
conditions. Two features recur across these data sets: the reduction grows
with frequency over the band where interaction noise dominates, and above it
the permeable surface generates self-noise that eventually cancels the benefit,
as it does for permeable and other modified trailing edges \citep{lee2021pas}.

The theory has advanced along two lines. The first treats the edge as a
scattering problem. \citet{jaworski2013} solved the poroelastic half-plane,
\citet{ayton2016} a rigid plate with a poroelastic extension, and
\citet{ayton2021} a chordwise-varying porosity, showing that the distribution
of permeability, not only its mean, controls the reduction. The second treats
the aerofoil as a finite chord. \citet{hajian2017} and \citet{baddoo2021}
solved the steady and unsteady incompressible problems for an arbitrary
chordwise porosity with a shed wake and the Kutta condition, and showed that
a permeable edge carries a loading singularity whose exponent depends on the
local permeability. The finite-chord problem has recently been carried into
compressible subsonic flow \citep{lee2026}, where the chord need not be
acoustically compact and the radiated field follows from the solution.
\citet{hales2024} combined a semi-infinite scattering solution with a
three-dimensional anisotropic turbulence model and predicted absolute spectra
for their perforated inserts.

What these treatments share is that the step from a measured material to the
boundary condition is where most of the residual error lies, and it is usually
closed by adjustment. \citet{hales2024} assembled the impedance of a perforated
sheet from the Rayleigh conductivity of an aperture \citep{rayleigh1870}, the
grazing-flow correction of \citet{howe1996}, the thickness correction of
\citet{luong2005} and the reactance of \citet{jing2000}; they found that this
impedance over-predicted the radiated noise at all frequencies, and introduced
a scaling function with three empirical constants tuned to one perforate at
one flow condition. The interpretation of the perforated-plate measurements of
\citet{chaitanya2020jsv} similarly rests on an assumed convection velocity. The impedance of perforates in grazing flow is well developed
in the acoustic-liner literature \citep{guess1975,melling1973,motsinger1991,
yu2008}, but its models are calibrated on thick turbulent boundary layers over
apertures much smaller than the layer. A leading-edge insert sits where the
boundary layer is thinnest, usually laminar, and comparable in thickness to
the apertures; it is not obvious that those models carry over.

A second question concerns compressibility. Leading-edge noise experiments
are typically carried out at Mach numbers of $0.05$--$0.2$, and incompressible
theories are commonly used for them. The quantity that controls acoustic
compactness, however, is the chord in acoustic wavelengths, not the Mach
number. At the frequencies where porous leading edges are effective, a chord of
a few hundred millimetres spans one or more acoustic wavelengths even at low
speed, so the loading radiates with retarded phase across the chord. Whether
this affects the predicted reduction, as distinct from the absolute level, has
not been tested against measurements.

This paper addresses both questions with a single calculation that contains
no constant adjusted to acoustic data. Its contributions are as follows. First,
the admittance of a perforated leading-edge insert is derived in full from the
Rayleigh conductivity of one aperture, with three corrections that are
computed rather than chosen: the plate thickness, the interaction between
neighbouring apertures through the function of \citet{fok1941}, and Howe's
grazing-flow blockage evaluated at the velocity averaged over the aperture
opening from the local boundary-layer profile. A bulk porous insert is
represented by its measured flow resistance and its pore-fluid inertia.
Second, these material laws are coupled to the compressible finite-chord
scattering solution of \citet{lee2026} and closed to the far field with the
finite span of the aerofoil and the aperture of the measuring array
represented. Third, the prediction is compared with the three data sets
described above, for perforated and bulk porous inserts, and the contributions
of each material term, of the observable and of compressibility are separated
by holding everything else fixed. Fourth, the comparison identifies where the
model fails, and why. Finally, a parameter study, made with the validated model,
isolates the geometric and flow parameters that control the reduction.

Section~\ref{sec:model} sets out the model: the scattering problem, the
material laws and the radiated field, with a verification of the radiation
calculation. Section~\ref{sec:experiments} compares the prediction with the
three experiments and identifies where it fails. Section~\ref{sec:param} uses
the model to isolate the parameters that control the reduction and discusses
its limits, and Section~\ref{sec:conclusions} concludes.

\section{Model}
\label{sec:model}

The model has three parts: the scattering problem for a permeable finite-chord
plate (Section~\ref{sec:formulation}), the material laws that supply its boundary
condition (Section~\ref{sec:material}), and the radiated field together with the
observables used in the comparisons and the numerical checks
(Section~\ref{sec:farfield}).

\subsection{The scattering problem}
\label{sec:formulation}

This subsection summarises the finite-chord formulation. Its derivation, the edge
analysis and the verification of the unsteady loading are given in
\citet{lee2026}; only what is needed to define the model and its inputs is
repeated here.

\subsubsection{Configuration}

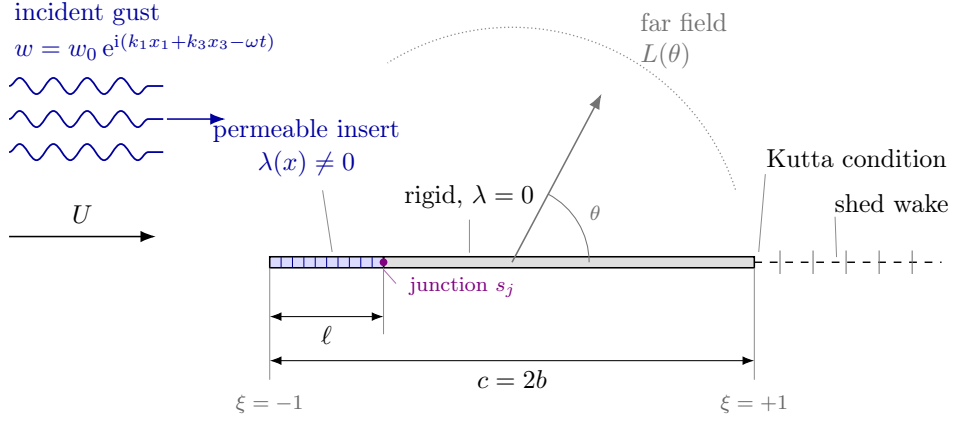
\begin{figure}[t]
\centering
\resizebox{0.92\textwidth}{!}{%
\begin{tikzpicture}[
  x=1cm,y=1cm,font=\footnotesize,
  gust/.style={draw=blue!62!black,line width=0.7pt,
               decorate,decoration={snake,amplitude=1.1mm,segment length=4.6mm}},
  ar/.style={-{Latex[length=2.1mm]},line width=0.6pt},
  dim/.style={{Latex[length=1.6mm]}-{Latex[length=1.6mm]},line width=0.4pt},
  lead/.style={draw=black!45,line width=0.3pt}]
\def\LE{0}\def\SJ{1.55}\def\TE{6.6}\def\TH{0.075}
\coordinate (O) at (3.30,0);
\fill[black!12] (\SJ,-\TH) rectangle (\TE,\TH);
\draw[line width=0.6pt] (\SJ,-\TH) rectangle (\TE,\TH);
\fill[blue!14] (\LE,-\TH) rectangle (\SJ,\TH);
\draw[line width=0.6pt] (\LE,-\TH) rectangle (\SJ,\TH);
\foreach \i in {1,...,9}{%
  \draw[blue!55!black,line width=0.4pt]
    ({\LE+\i*(\SJ-\LE)/10},-\TH) -- ({\LE+\i*(\SJ-\LE)/10},\TH);}
\fill[violet] (\SJ,0) circle (1.5pt);
\draw[dashed,line width=0.55pt] (\TE,0) -- (9.15,0);
\foreach \x in {6.95,7.4,7.85,8.3,8.75}{\draw[black!55,line width=0.4pt] (\x,-0.16) -- (\x,0.16);}
\node[anchor=west] at (6.62,1.42) {Kutta condition};
\draw[lead] (6.75,1.20) -- (6.66,0.14);
\node[anchor=west] at (7.55,0.80) {shed wake};
\draw[lead] (7.70,0.62) -- (7.72,0.07);
\foreach \y in {1.50,1.95,2.40}{\draw[gust] (-3.55,\y) -- (-1.45,\y);}
\draw[ar,draw=blue!62!black] (-1.40,1.95) -- (-0.60,1.95);
\node[anchor=west,text=blue!62!black,align=left] at (-3.62,3.15)
  {incident gust\\[2pt] $w=w_0\,\E^{\I(k_1x_1+k_3x_3-\omega t)}$};
\draw[ar] (-3.55,0.35) -- (-1.55,0.35);
\node[above] at (-2.55,0.39) {$U$};
\draw[black!45,line width=0.3pt] (\LE,-\TH) -- (\LE,-1.60);
\draw[black!45,line width=0.3pt] (\SJ,-\TH) -- (\SJ,-0.98);
\draw[black!45,line width=0.3pt] (\TE,-\TH) -- (\TE,-1.60);
\draw[dim] (\LE,-0.74) -- (\SJ,-0.74);
\node[below] at ({(\LE+\SJ)/2},-0.72) {$\ell$};
\draw[dim] (\LE,-1.34) -- (\TE,-1.34);
\node[below] at ({(\LE+\TE)/2},-1.32) {$c=2b$};
\node[below,font=\scriptsize,text=black!60] at (\LE,-1.66) {$\xi=-1$};
\node[below,font=\scriptsize,text=black!60] at (\TE,-1.66) {$\xi=+1$};
\node[text=blue!62!black,align=center,anchor=south] at (0.50,1.02)
  {permeable insert\\[1pt] $\lambda(x)\neq0$};
\draw[lead] (0.72,0.98) -- (0.85,0.14);
\node[anchor=south] at (2.72,0.60) {rigid, $\lambda=0$};
\draw[lead] (2.72,0.58) -- (2.72,0.13);
\node[text=violet,anchor=west,font=\scriptsize] at (1.76,-0.34) {junction $s_j$};
\draw[lead,draw=violet!60] (1.72,-0.29) -- (\SJ,-0.10);
\draw[black!45,densely dotted,line width=0.5pt] (O) ++(18:3.20) arc (18:122:3.20);
\draw[ar,draw=black!55] (O) -- ($(O)+(62:2.60)$);
\draw[black!55,line width=0.4pt] (O) ++(0:1.05) arc (0:62:1.05);
\node[text=black!55,font=\scriptsize] at ($(O)+(31:1.32)$) {$\theta$};
\node[text=black!55,anchor=west,align=left] at ($(O)+(62:3.42)$) {far field\\ $L(\theta)$};
\end{tikzpicture}}
\caption{The configuration. A flat plate of chord $c = 2b$ carries a permeable
insert of length $\ell$ at the leading edge, across which the admittance group
is $\lambda(x)$; the remainder is rigid. A frozen gust convects at the mean flow
speed $U$. The solution carries a shed wake with the unsteady Kutta condition at
the trailing edge, and the far-field angle $\theta$ is measured from the
downstream chord line.}
\label{fig:geom}
\end{figure}

A thin flat plate of chord $c = 2b$ lies along $-b \le x_1 \le b$, $x_2 = 0$,
in a uniform stream $U$ of density $\rho_0$ and sound speed $c_0$, at Mach number
$M = U/c_0$ with $\beta_M = \sqrt{1-M^2}$ (Fig.~\ref{fig:geom}). A permeable
insert occupies the leading-edge region $0 \le x \le \ell$, where
$x = x_1 + b$ is the distance from the leading edge, and the remainder of the
chord is rigid. A frozen gust of upwash $w = w_0\exp[\I(k_1x_1 + k_3x_3 -
\omega t)]$, with $k_1 = \omega/U$, is incident. The time dependence
$\E^{-\I\omega t}$ is used throughout. The Prandtl--Glauert transformation maps
the problem to a Helmholtz problem on a plate of stretched half-length
$b/\beta_M$; with the chordwise coordinate normalised on that length,
$\xi = x_1/b \in [-1,1]$, the two reduced wavenumbers are
\begin{equation}
k_e = \frac{k_0 b}{\beta_M^2}, \qquad \kb = \frac{k_1 b}{\beta_M^2},
\qquad k_e = M\kb ,
\label{eq:nondim}
\end{equation}
with $k_0 = \omega/c_0$. The first measures the chord in acoustic wavelengths and
the second in hydrodynamic wavelengths; the classical reduced frequency is
$\sigma = \omega b/U = \beta_M^2\kb$. Because $k_e = M\kb$, a chord can be
acoustically non-compact at low Mach number once $\kb$ is large, and in the
comparisons of Section~\ref{sec:hales} $\kb$ reaches $135$.

\subsubsection{The permeable boundary condition}
\label{sec:bc}

The insert is modelled as a homogenised sheet across which the pressure jump
$\Delta p$, upper surface minus lower, drives a seepage velocity $v_s$, positive
in the $x_2$ direction, through the normalised transfer admittance
$\Ad = 1/\Zt$,
\begin{equation}
v_s = -\frac{\Ad\,\Delta p}{\rho_0 c_0},
\label{eq:vs}
\end{equation}
so that fluid passes from the side at higher pressure to the side at lower
pressure. The linearised momentum relation across the sheet,
$\Delta p = -\rho_0(\partial_t + U\partial_{x_1})[\varphi]$, with $[\varphi]$ the
jump in disturbance potential, and continuity of the normal velocity, which
requires the imposed upwash $w$ plus the disturbance velocity to equal the
seepage velocity on both faces, together give
\begin{equation}
\frac{\partial\varphi}{\partial x_2}\bigg|_{x_2=0}
 = -w + v_s
 = -w + \frac{\Ad}{c_0}
   \left(\frac{\partial}{\partial t} + U\frac{\partial}{\partial x_1}\right)
   [\varphi] .
\label{eq:bcconv}
\end{equation}
The seepage is proportional to the rate of change of the potential jump
following the mean flow. After non-dimensionalisation the material enters the
problem through the single group
\begin{equation}
\lambda = \frac{2\Ad M}{\beta_M}
        = \frac{2\rho_0 U}{\beta_M Z_d},
\label{eq:lambda}
\end{equation}
where $Z_d = \rho_0c_0\Zt = -\Delta p/v_s$ is the dimensional transfer
impedance. A rigid surface
has $\lambda = 0$ and a fully transparent one $\lambda\to\infty$. Everything
specific to the material, including the grazing flow over it, is therefore
contained in $Z_d$, which is the subject of Section~\ref{sec:material}.

\subsubsection{Integral equation and edge behaviour}
\label{sec:sie}

Representing the plate and its wake by a sheet of pressure dipoles and
enforcing Eq.~\eqref{eq:bcconv} gives, after the Prandtl--Glauert transformation, a
singular integral equation for the normalised pressure jump $\Pi(\xi)$,
\begin{equation}
\lambda(\xi)\,\Pi(\xi) + \frac{1}{\pi}\pvint_{-1}^{1}\frac{\Pi(s)}{\xi-s}\,\D s
+ \int_{-1}^{1} \mathcal{K}(\xi,s)\,\Pi(s)\,\D s = \Wup(\xi)
 = -\E^{\I(\kb\xi + \sigma)} ,
\label{eq:sie}
\end{equation}
in which the Cauchy integral is the singular part of the compressible
thin-aerofoil kernel of \citet{possio1938} and $\mathcal{K}$ is a bounded
remainder containing the Hankel-function terms that carry the acoustic
wavenumber $k_e$. The forcing is the physical upwash, normalised on $w_0$ and
referred to the leading edge, $w = \E^{\I\sigma(\xi+1)}$, multiplied by the
Prandtl--Glauert phase $\E^{\I Mk_e\xi}$; since $\sigma + Mk_e = \kb$, the gust
convects at $\kb$ in the transformed variables. The same phase appears, inverted,
in the output: the physical pressure jump, upper surface minus lower, is
$\Pi(\xi)\,\E^{-\I Mk_e\xi}$ \citep{lee2026}. Kelvin's
theorem and the unsteady Kutta condition close the problem. Setting $k_e\to0$
removes the Hankel terms and recovers the incompressible equation of
\citet{baddoo2021}.

Near the leading edge the bounded remainder is subdominant, and a
Riemann--Hilbert argument gives the local behaviour
$\Pi \sim (1+\xi)^{-\beta}$ with
\begin{equation}
\beta = \frac{1}{\pi}\cot^{-1}\lambda(-1) ,
\label{eq:beta}
\end{equation}
which falls from the rigid value $\tfrac12$ as the edge becomes more permeable.
At the junction $s_j = 2\ell/c - 1$, where $\lambda$ drops to zero, the loading
carries a weaker singularity $|\xi - s_j|^{-\delta}$ with
$\delta = \tfrac12 - \pi^{-1}\cot^{-1}\lambda(s_j^-)$, the exponent
of Eq.~\eqref{eq:beta} evaluated just upstream of the junction. These are the incompressible exponents of
\citet{baddoo2021}; they survive compressibility unchanged in form because the
non-Cauchy part of the kernel is bounded \citep{lee2026}. The equation is
solved by collocation in a basis of weighted Jacobi polynomials that carry the
exponent of Eq.~\eqref{eq:beta} at the leading edge and the rigid exponent at the
trailing edge, supplemented by two junction modes carrying $\delta$; the Kutta
condition is built into the basis, and Kelvin's theorem supplies one extra
row. A variable $\lambda(\xi)$ enters only on the diagonal of the collocation
matrix, so the chordwise variation of the admittance derived in
Section~\ref{sec:material} costs nothing. The mode count and the checks made for the
present calculations are given in Section~\ref{sec:numerics}.

\subsubsection{Similarity}
\label{sec:similarity}

Writing $\mu = -\I k_0\Ad$ for the porosity parameter used by
\citet{jaworski2013} and
using $k_0 = Mk_1$, Eq.~\eqref{eq:lambda} becomes
\begin{equation}
\lambda = \frac{2\I\mu}{\beta_M k_1} = \frac{\I\mu U}{\pi\beta_M f} .
\label{eq:lamsim}
\end{equation}
The permeability a gust sees is the porosity parameter measured against the
hydrodynamic wavenumber, not the acoustic one. The two material classes
considered here then behave oppositely. A purely reactive perforate has
$\Zt \propto k_0$, so $\mu$ is a geometric inverse length and $|\lambda|$ falls
as $U/f$. A purely resistive layer has $Z_d$ independent of frequency, so
$\lambda = 2\rho_0U/(\beta_M Z_d)$ is constant in frequency and grows linearly
with $U$. Real perforates in grazing flow lie between these limits, and where
they lie is decided by the aperture Strouhal number (Section~\ref{sec:lowfreq}).

\subsection{Material laws}
\label{sec:material}

The scattering problem needs one input from the material, the transfer
impedance $Z_d$ in Eq.~\eqref{eq:lambda}, as a function of position along the
insert and of frequency. This subsection derives it for a perforated sheet in
grazing flow and for a bulk porous layer, from measured geometry or measured
permeability, without reference to acoustic data. The three perforated
inserts of \citet{hales2024}, which are compared with measurements in
Section~\ref{sec:hales}, serve as the running example. Labelled as in their paper,
they are square arrays of circular holes in a plate $1.5$\,mm thick: case~20
has hole radius $R = 0.5$\,mm at pitch $P = 3.0$\,mm, case~30 has
$R = 1.0$\,mm and $P = 4.0$\,mm, and case~40 has $R = 0.5$\,mm and
$P = 1.5$\,mm, so that the holes occupy $8.7$, $19.6$ and $34.9\,\%$ of the
surface.

\subsubsection{The Rayleigh conductivity of one aperture}
\label{sec:KR}

Consider one circular aperture of radius $R$ in an otherwise rigid wall, driven
by a harmonic pressure difference $\Delta p$ measured far enough from the
aperture that its near field has decayed, and let $Q$ be the volume flux through
it from the upper to the lower face, the direction in which a positive $\Delta p$
drives it. The Rayleigh conductivity is defined by
\citep{rayleigh1870,howe1998}
\begin{equation}
K_R \equiv \frac{-\I\omega\rho_0\,Q}{\Delta p} .
\label{eq:KRdef}
\end{equation}
It has the dimensions of length and depends on the aperture alone. For a
circular aperture in a wall of zero thickness without flow, $K_R = 2R$. Three
effects modify this value, and each can be computed.

\paragraph{Wall thickness} An aperture in a wall of thickness $h$ is a short
channel. The fluid in it has inertia $\rho_0 h$ per unit area, and the
zero-thickness conductivity $2R$ corresponds to an effective length
$\pi R^2/(2R) = \pi R/2$, that is, two end corrections of $\pi R/4$, one at
each mouth. Adding the channel to the two end corrections in series gives
\citep{luong2005,jing2000}
\begin{equation}
K_R = \frac{\pi R^2}{h + \tfrac12\pi R} = \frac{2R}{1 + 2h/(\pi R)} .
\label{eq:KRthick}
\end{equation}
For the perforates of \citet{hales2024}, $2h/(\pi R) = 0.95$--$1.91$, so the
thickness alone halves the conductivity or more.

\paragraph{Aperture interaction} When apertures are close together, the
reactive near field that the end correction represents is shared with the
neighbours and each end correction is reduced. For a regular array of holes of
diameter $d = 2R$ at pitch $P$, the reduction is given by the interaction
function of \citet{fok1941},
\begin{multline}
\psi(\varsigma) = 1 - 1.4092\,\varsigma + 0.33818\,\varsigma^3
              + 0.06793\,\varsigma^5 \\
              - 0.02287\,\varsigma^6
              + 0.03015\,\varsigma^7 - 0.01641\,\varsigma^8,
\qquad \varsigma = \frac{d}{P},
\label{eq:fok}
\end{multline}
which falls from $\psi(0) = 1$ for an isolated hole towards zero as neighbouring
holes touch \citep{ingard1953,melling1973,guess1975}. Each end correction is
multiplied by $\psi$, so that $K_R = \pi R^2/(h + \tfrac12\pi R\psi)$. Cases~20,
30 and 40 have $\varsigma = 0.33$, $0.50$ and $0.67$ and
therefore $\psi = 0.543$, $0.340$ and $0.169$ (Fig.~\ref{fig:material}($a$)):
for the most open of them each end correction is $17\,\%$ of its isolated value.
Leading-edge inserts are usually close-packed, because a large open area is
needed for a useful admittance, and $\psi$ is then a leading-order term rather
than a refinement.

\paragraph{Grazing flow} A mean flow grazing the aperture sheds vorticity from
the upstream rim, and the resulting shear layer impedes the flux.
\citet{howe1996} modelled the shear layer as a vortex sheet convected across the
aperture at a velocity $U_c$ and obtained the conductivity in grazing flow as the
no-flow value multiplied by a complex blockage factor $\Hgr$,
\begin{equation}
\Hgr(\Sr) = 1 + \frac{\tfrac{\pi}{2}I_1(\Sr)\,\E^{-\Sr} - \I K_1(\Sr)\sinh\Sr}
                    {\Sr\left[\tfrac{\pi}{2}I_1(\Sr)\,\E^{-\Sr} + \I K_1(\Sr)\cosh\Sr\right]},
\qquad \Sr = \frac{\omega R}{U_c},
\label{eq:howe}
\end{equation}
with $I_1$ and $K_1$ modified Bessel functions. $\Hgr$ depends on the aperture
Strouhal number alone. It tends to unity as $\Sr\to\infty$, where the vortex
sheet is too slow to matter, and to $-\I\pi\Sr/4$ as $\Sr\to0$
(Fig.~\ref{fig:material}($b$)).

Collecting the three effects, and applying the blockage to the whole
conductivity as \citet{luong2005} and \citet{hales2024} do,
\begin{equation}
K_R = \frac{2R\,\Hgr(\Sr)}{\Dc}, \qquad \Dc = \psi(\varsigma) + \frac{2h}{\pi R} .
\label{eq:KR}
\end{equation}
Equation~\eqref{eq:KR} reduces to Eq.~\eqref{eq:KRthick} when $\psi = \Hgr = 1$ and
to Rayleigh's $2R$ when in addition $h = 0$. The alternative placement, with
the blockage applied to the end corrections only and the channel inertia left in
series, is tested against data in Section~\ref{sec:isolation}.

\subsubsection{From one aperture to a sheet}

Let $\alpha_H$ be the open-area ratio of the sheet, the fraction of its area
occupied by the apertures, and $N_a$ the number of holes per unit area of the
sheet. Since each aperture has area $\pi R^2$, $N_a = \alpha_H/(\pi R^2)$; for a
square array of pitch $P$, $\alpha_H = \pi R^2/P^2$ and $N_a = 1/P^2$. Each aperture passes $Q = K_R\Delta p/(-\I\omega\rho_0)$ by
Eq.~\eqref{eq:KRdef}, so the mean seepage velocity, positive in the $x_2$ direction
as in Eq.~\eqref{eq:vs}, is $v_s = -N_aQ$, and the transfer impedance normalised on
$\rho_0c_0$ is
\begin{equation}
\Zt = -\frac{\Delta p}{\rho_0c_0v_s} = -\frac{\I k_0}{N_aK_R} - \I k_0h .
\label{eq:Zperf}
\end{equation}
The first term is the aperture contribution. The second is the inertia of a
fluid layer of the sheet thickness, which appears in the impedance of a thick
perforated plate \citep{jing2000}. It is small here, $1$--$6\,\%$ of the first
term at $500$\,Hz and $4$--$21\,\%$ at $2$\,kHz for the three perforates of
\citet{hales2024}, but not negligible for the most open one at the top of the
band, and it is retained. Equations~\eqref{eq:KR} and \eqref{eq:Zperf}, with
Eq.~\eqref{eq:lambda}, are the complete route from a stated perforate geometry to the
coefficient of Eq.~\eqref{eq:sie}.

\subsubsection{The low-frequency limit}
\label{sec:lowfreq}

Because $\Hgr\to-\I\pi\Sr/4$ as $\Sr\to0$, the conductivity of Eq.~\eqref{eq:KR}
vanishes linearly in $\omega$. It does not follow that the apertures close.
The factor $k_0$ in Eq.~\eqref{eq:Zperf} vanishes at the same rate, and with
$N_aK_R \to -\I\alpha_H\omega/(2U_c\Dc)$,
\begin{equation}
\Zt \;\longrightarrow\; \frac{2U_c\Dc}{c_0\,\alpha_H},
\qquad
\lambda \;\longrightarrow\; \lambda_0
 = \frac{\alpha_H}{\beta_M\,\Dc}\,\frac{U}{U_c} .
\label{eq:lam0}
\end{equation}
Both are real and independent of frequency. At low aperture Strouhal number the
grazing flow does not block a perforate; it makes it resistive, with a
resistance set by the shed vorticity, and the surface keeps a
frequency-independent permeability. The perforate then behaves as the resistive
class of Section~\ref{sec:similarity}, and it becomes reactive as $\Sr$ rises through
unity. For the inserts of \citet{hales2024} at $U = 14.8$\,m\,s$^{-1}$,
Eq.~\eqref{eq:lam0} gives insert-averaged values $\lambda_0 = 0.07$, $0.21$ and
$0.35$ for cases~20, 30 and 40, so that $\beta = 0.48$, $0.43$ and $0.39$ at the low-frequency end:
modest departures from the rigid $\tfrac12$ that nonetheless produce reductions
of several decibels at higher $\kb$ (Section~\ref{sec:hales}).
Equation~\eqref{eq:lam0} also identifies the parameters that a design can
change: the open area $\alpha_H$ in the numerator, and in the denominator the
thickness-to-radius ratio and the interaction function through $\Dc$ and the
boundary layer through $U_c/U$. The parameter study of Section~\ref{sec:param} tests
whether $\lambda_0$ organises the computed reductions.

\subsubsection{The aperture-averaged grazing velocity}
\label{sec:bl}

\begin{figure}[t]
\centering
\includegraphics[width=\textwidth]{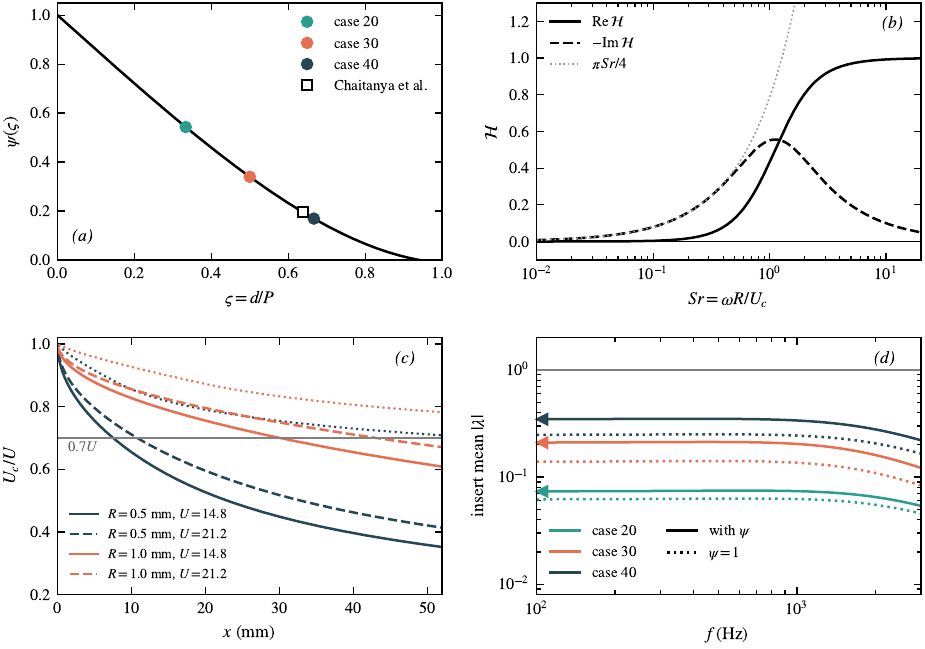}
\caption{Ingredients of the perforate admittance. ($a$) Fok's interaction
function of Eq.~\eqref{eq:fok}, with the perforates of \citet{hales2024} (circles)
and \citet{chaitanya2020jsv} (square). ($b$) Howe's blockage factor
of Eq.~\eqref{eq:howe}, real and imaginary parts, with the low-Strouhal asymptote
$\pi\Sr/4$. ($c$) The aperture-averaged grazing velocity of Eq.~\eqref{eq:uc} along a
$52$\,mm insert for two hole radii and two speeds; dotted, the turbulent closure
of Eq.~\eqref{eq:ucturb}; horizontal line, the conventional $0.7U$. ($d$) The
insert-averaged $|\lambda|$ of the three perforates of \citet{hales2024} at
$U = 14.8$\,m\,s$^{-1}$, with (solid) and without (dotted) Fok's function;
the markers on the left axis are the low-frequency limits of Eq.~\eqref{eq:lam0}.}
\label{fig:material}
\end{figure}

Howe's correction enters through $\Sr = \omega R/U_c$, and his derivation
assumes that the aperture sits in a uniform grazing stream. At a leading-edge
insert it does not. For the inserts of \citet{hales2024} the local Reynolds
number is $\Rey_x \le 7.4\times10^4$, the layer is laminar and its thickness is
comparable to the hole radius, so the aperture spans the shear layer. The
velocity that convects the rim vorticity across the aperture is therefore taken
as the average over the aperture height,
\begin{equation}
U_c(x) = \frac{1}{R}\int_0^R u(y;x)\,\D y ,
\label{eq:ucdef}
\end{equation}
where $u(y;x)$ is the local mean-velocity profile. For a Blasius layer
\citep{blasius1908,schlichting2017}, $u/U = F'(\eta)$ with
$\eta = y\sqrt{U/(\nu x)}$, and the integral collapses to
\begin{equation}
\frac{U_c}{U} = \frac{1}{\eta_R}\int_0^{\eta_R}F'(\eta)\,\D\eta
             = \frac{F(\eta_R)}{\eta_R}, \qquad
\eta_R = R\sqrt{\frac{U}{\nu x}} ,
\label{eq:uc}
\end{equation}
with $F$ the Blasius stream function. For an aperture much larger than the
local layer, $F(\eta)\to\eta - 1.7208$ gives $U_c/U \to 1 - 1.7208/\eta_R$; for
one much smaller, $U_c/U \to F''(0)\eta_R/2 = 0.166\,\eta_R$. Along a $52$\,mm
insert with $R = 0.5$\,mm at $U = 14.8$\,m\,s$^{-1}$, $U_c/U$ falls from $0.97$
at the nose to $0.35$ at the junction, with a mean of $0.52$; for $R = 1.0$\,mm
the mean is $0.74$ (Fig.~\ref{fig:material}($c$)). The value $0.7$ commonly
assumed is therefore neither the local value nor the average, and the hole
radius enters the blockage through $\eta_R$ as well as through $\Sr$.

Equation~\eqref{eq:ucdef} makes no assumption about the state of the layer; only
Eq.~\eqref{eq:uc} is laminar. For a turbulent layer with a $1/7$-power profile and
$\delta_{99} = 0.37\,x\,\Rey_x^{-1/5}$ the same average is again closed-form,
\begin{equation}
\frac{U_c}{U} =
\begin{cases}
\dfrac{7}{8}\left(\dfrac{R}{\delta_{99}}\right)^{1/7}, & R \le \delta_{99},\\[8pt]
1 - \dfrac{\delta_{99}}{8R}, & R > \delta_{99}.
\end{cases}
\label{eq:ucturb}
\end{equation}
Over the inserts of \citet{hales2024} the turbulent average is $0.79$--$0.87$,
close to the conventional constant, while the laminar one is $0.52$--$0.78$.
The conventional value therefore reflects the developed turbulent layers in
which perforates are usually characterised, and it overstates $U_c$ at a
leading-edge insert. The closure is selected from the flow, not from the noise
data: at the junction of the inserts of \citet{hales2024} $\Rey_x$ is
$5.1$--$7.4\times10^4$, an order of magnitude below natural transition, and the
roughness Reynolds number of the perforation, $u_\tau d/\nu$ with $u_\tau$ from
the laminar wall shear at the junction, is $38$--$99$, well below the values of
several hundred needed to trip a laminar layer \citep{schlichting2017}. The
turbulent closure is tested against the data in Section~\ref{sec:isolation}.

\subsubsection{A bulk porous layer}
\label{sec:bulk}

For a rigid-frame porous material the filtration velocity $v_s$ obeys, at low
frequency, a momentum balance of Darcy's resistance and the inertia of the pore
fluid,
\begin{equation}
-\frac{\partial p}{\partial x_2} = \frac{\mu_{\rm air}}{\kappa}\,v_s
  + \frac{\rho_0\alpha_\infty}{\phi}\,\frac{\partial v_s}{\partial t} ,
\label{eq:darcy}
\end{equation}
with $\kappa$ the permeability, $\phi$ the porosity, $\alpha_\infty$ the
tortuosity and $\mu_{\rm air}$ the dynamic viscosity of air. Integrated across a
layer of thickness $h$ whose compressibility is negligible, it gives the
transfer impedance
\begin{equation}
Z_d = r\,h - \I\omega\rho_0\frac{\alpha_\infty}{\phi}\,h, \qquad
r = \frac{\mu_{\rm air}}{\kappa},
\label{eq:Zbulk}
\end{equation}
and hence $\lambda = 2\rho_0U/(\beta_MZ_d)$ from Eq.~\eqref{eq:lambda}. Equation
\eqref{eq:Zbulk} is the low-frequency limit of the equivalent-fluid model of
Johnson, Champoux and Allard
\citep{johnson1987,champoux1991,allard2009}, which is used in Section~\ref{sec:bowen}
as a check. The first term is the resistive limit of Section~\ref{sec:similarity};
the second makes the layer reactive above the frequency
$\phi r/(2\pi\rho_0\alpha_\infty)$, which for the materials of
Section~\ref{sec:bowen} is a few hundred hertz, so the inertia is not a correction
there. On an aerofoil the seepage path through the nose varies along the chord;
it is taken as the local thickness of the section, bounded below by the
leading-edge diameter, so that $\lambda$ again varies along the insert.

\subsection{Far field, observables and verification}
\label{sec:farfield}

\subsubsection{Radiation from the loading}

The turbulence is represented as a superposition of frozen gusts; at a
given frequency they all have $k_1 = \omega/U$ and differ in their spanwise
wavenumber $k_3$. For a gust of spanwise wavenumber $k_3$ the far-field pressure radiated by the
plate and its wake follows from the loading through the convected free-field
Green's function \citep{amiet1975}. At an observer at distance $\mathcal{R}$ in
the plane $x_3 = 0$, at angle $\theta$ from the downstream chord line, the
chordwise part of the radiation is carried by the transfer function
\begin{equation}
L(\theta;k_3) = \frac{1}{\pi\beta_M}\int_{-1}^{1}\Pi(\xi;k_3)\,\E^{-\I Mk_e\xi}\,
\exp\!\Big[\I k_e\Big(M - \frac{x_1}{S_0}\Big)\xi\Big]\D\xi ,
\label{eq:L}
\end{equation}
with $S_0^2 = x_1^2 + \beta_M^2x_2^2$. This is Amiet's aeroacoustic transfer function with his rigid semi-infinite
loading replaced by the physical pressure jump $\Pi\,\E^{-\I Mk_e\xi}$ of the
finite-chord porous solution of Eq.~\eqref{eq:sie}. With the normalisation of $\Pi$
in Eq.~\eqref{eq:sie}, $|L|$ tends to $|S(\sigma)|/\beta_M$, with $S$ the Sears
function, for a rigid plate whose chord is acoustically compact. An oblique gust
leaves $\kb$ and the Prandtl--Glauert phase unchanged and replaces the acoustic
wavenumber in the kernel by
\begin{equation}
\tilde{k}_e(k_3) = \sqrt{k_e^2 - (k_3b/\beta_M)^2}, \qquad
\operatorname{Im}\tilde{k}_e \ge 0 ,
\label{eq:kappa}
\end{equation}
which becomes imaginary, and the response evanescent, beyond the spanwise
cut-off $k_3 = k_0/\beta_M$ \citep{lee2026}.

For a plate of span $2d$ in turbulence with upwash spectrum
$\Phi_{ww}(k_1,k_3)$, summing the gust components incoherently gives the
far-field spectrum at an observer in the mid-span plane as
\begin{equation}
S_{pp}(\theta,\omega) = C(\mathcal{R},\theta,\omega)\int_{-\infty}^{\infty}
\Phi_{ww}(k_1,k_3)\,\big|L(\theta;k_3)\big|^2\,
\frac{\sin^2(k_3d)}{k_3^2}\,\D k_3 ,
\label{eq:spp}
\end{equation}
in which $C$ collects the observer distance, the density and the other factors
that do not depend on the material \citep{amiet1975,roger2010}. As $d\to\infty$
the span factor selects $k_3 = 0$ and Eq.~\eqref{eq:spp} reduces to Amiet's
large-span form. The turbulence intensity, the observer distance and every
factor in $C$ cancel from a reduction; the spectrum $\Phi_{ww}$ and the span do
not cancel in general, because they weight the $k_3$ dependence of the rigid and
permeable loadings differently. Here $\Phi_{ww}$ is the von K\'arm\'an spectrum
with the measured integral length scale.

\subsubsection{Observables}
\label{sec:observables}

Every comparison below is of a reduction, rigid minus permeable, and the
reduction depends on what is observed. Writing $\langle\,\cdot\,\rangle$ for an
average over $\theta$ with weight $\sin\theta$, over the window
$|\theta - 90^\circ| \le 20^\circ$ unless another arc is stated, four
observables are used:
\begin{align}
\dSPL_B &= 10\log_{10}\frac{|L_{\rm rigid}(90^\circ;0)|^2}{|L_{\rm perm}(90^\circ;0)|^2},
&
\dSPL_W &= 10\log_{10}\frac{\langle|L_{\rm rigid}(\theta;0)|^2\rangle}{\langle|L_{\rm perm}(\theta;0)|^2\rangle},
\label{eq:obsBW}\\
\dSPL_S &= 10\log_{10}\frac{\langle S_{pp,\rm rigid}\rangle}{\langle S_{pp,\rm perm}\rangle},
&
\dPWL &= 10\log_{10}\frac{\langle|L_{\rm rigid}(\theta;0)|^2\rangle_{\rm arc}}{\langle|L_{\rm perm}(\theta;0)|^2\rangle_{\rm arc}} .
\label{eq:obsSP}
\end{align}
Observable B is a single gust, the spanwise-uniform component $k_3 = 0$,
from which the turbulence spectrum cancels entirely, at a single observer at
broadside:
the large-span limit at one angle, and the idealised limit of what follows.
Observable W keeps the single gust and averages over the angular window.
Observable S, the span-and-array observable, evaluates Eq.~\eqref{eq:spp} for the
rigid and permeable plates and averages over the same window. Observable P is a
power reduction over the arc of a measurement reported as sound power.
Observable S is the one that corresponds to a beamformed measurement: the
integration region of the source map fixes the span, and the array occupies a
range of angles. The half-width of $20^\circ$ is an assumption where the array
aperture is not reported, and its effect is quantified in Section~\ref{sec:hales}. For
an acoustically compact chord $|L|\propto$ const in $\theta$ for both surfaces
and the four observables coincide; at the frequencies of interest here they do
not, and the differences are part of the results.

\subsubsection{Numerical parameters}
\label{sec:numerics}

The mode count is set automatically as $N \simeq 1.6\kb + 2k_e + 12$
\citep{lee2026}. Three numerical points specific to the present calculations
were checked before any comparison with data. First, for the configuration of
Section~\ref{sec:hales} the automatic count is not sufficient in mid-band: at
$750$\,Hz the broadside reduction changes by up to $0.7$\,dB between the
automatic count and $1.5$ times it, whereas $1.5$, $2$ and $2.5$ times it
agree to within about $0.1$\,dB at $550$, $750$ and $1000$\,Hz. All
calculations for perforated inserts therefore use $1.5$ times the automatic
count. Second, when the junction exponent is small, $\delta \lesssim 0.15$, the
two junction modes become nearly linearly dependent on the polynomial basis once
$N$ exceeds about $390$, and the solution degrades abruptly; the count is
therefore capped at $360$. At the top of the band the reduction varies by
$\pm0.07$\,dB for $N$ between $200$ and $360$. Together with the agreement of
$1.5$, $2$ and $2.5$ times the automatic count, this places the numerical
uncertainty of the reductions at about a tenth of a decibel. The junction
modes themselves matter less than the mode count: switching them off changes
the reductions by at most $0.39$\,dB ($0.14$\,dB rms over $168$ cases). Third, the incompressible limit is computed with the same
solver at $k_e = 10^{-4}$ and $M = 0$; smaller values make the system
ill-conditioned at large $N$ (at $k_e = 10^{-6}$ the rigid lift is in error by
$1.6\,\%$ at $\kb = 22$ with three times the automatic mode count), whereas
$k_e = 10^{-4}$ reproduces the Sears function at every $N$ tested. For the bulk
porous aerofoil of Section~\ref{sec:bowen}, whose admittance varies continuously along
a thickening nose, three times the automatic mode count is used; doubling or
quadrupling it changes the reduction by about $0.3$\,dB.

\subsubsection{Verification of the radiated field}
\label{sec:verify}

\begin{figure}[t]
\centering
\includegraphics[width=\textwidth]{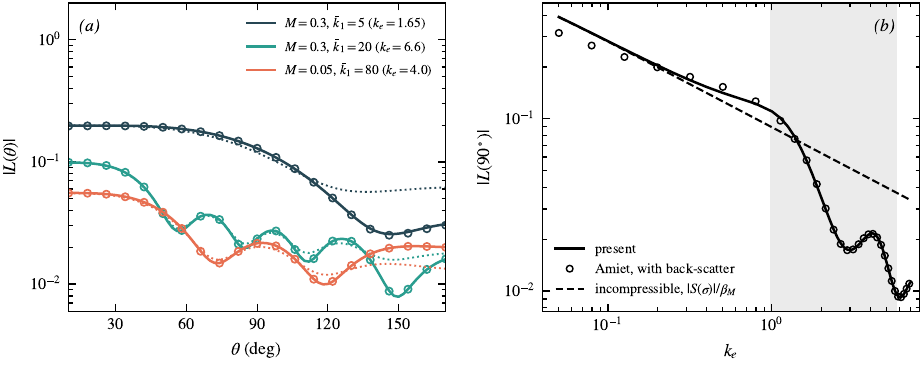}
\caption{Verification of the radiated field for a rigid plate. ($a$) Directivity
of $|L|$, Eq.~\eqref{eq:L}: lines, present solution; symbols, Amiet's leading-edge
solution with the trailing-edge back-scatter correction \citep{roger2010};
dotted, the leading-edge term alone. ($b$) Broadside $|L|$ at $M = 0.05$ against
$k_e$: present solution (solid), Amiet with back-scatter (symbols), and the
incompressible compact limit $|S(\sigma)|/\beta_M$ (dashed). The shaded band is the range of the comparisons with \citet{hales2024} in Section~\ref{sec:hales}.}
\label{fig:verify}
\end{figure}

The unsteady loading has been verified in \citet{lee2026} against the Sears and
Theodorsen functions, the incompressible porous solutions of
\citet{baddoo2021} computed with their published code, and an independent
compressible solution in physical variables. What remains to be checked here
is the radiation, Eq.~\eqref{eq:L}, at finite $k_e$. For a rigid plate in a
supercritical gust, \citet{amiet1975} gave a high-frequency approximation to
$L$ built from the leading-edge Schwarzschild solution, and a trailing-edge
back-scatter correction makes it accurate for $\bar\mu = Mk_1b/\beta_M^2 = k_e$
of order unity and above \citep{roger2010}. Figure~\ref{fig:verify}($a$)
compares Eq.~\eqref{eq:L} with that approximation. At $M = 0.3$ and $0.05$, over
$20^\circ \le \theta \le 160^\circ$, the two agree to within $0.35$\,dB for
$k_e \ge 1$ and $0.1$\,dB for $k_e \ge 4$; the leading-edge term alone,
without the back-scatter correction, is in error by up to $7$\,dB. At lower
$k_e$ the approximation is not expected to hold, and it departs by up to
$1.9$\,dB at $k_e = 0.05$ (Fig.~\ref{fig:verify}($b$)), while the present
solution tends to the Sears function, to $0.05$\,dB at $k_e \le 0.1$. The
calculation therefore reproduces both classical limits, each where it applies,
with no adjustable normalisation.

Figure~\ref{fig:verify}($b$) also shows why compressibility matters in the
experiments that follow. At $M = 0.05$, the incompressible broadside level,
which is the Sears function radiating from a compact source, departs from the
exact level once $k_e$ approaches unity: it is $1.8$\,dB lower at $k_e = 1$
and $5$--$12$\,dB higher over $k_e = 2$--$7$. The comparisons with
\citet{hales2024} in Section~\ref{sec:hales} span $k_e = 0.97$--$5.8$. This is a statement about the
rigid reference; its consequence for the reduction is examined in
Section~\ref{sec:compress}.

\section{Comparison with experiments}
\label{sec:experiments}

The prediction is compared with three experiments that together cover both
material classes: perforated leading-edge inserts (Section~\ref{sec:hales}), bulk
porous leading-edge inserts (Section~\ref{sec:bowen}), and perforated plates whose
measured spectra expose a limit of the model (Section~\ref{sec:chaitanya}). The
same solver and the same material laws are used throughout, and no input is
adjusted to the acoustic data.

\subsection{Perforated inserts: Hales et al.}
\label{sec:hales}

\subsubsection{The experiment and the inputs}

\citet{hales2024} measured perforated leading-edge inserts on a flat-plate
aerofoil in the UNSW anechoic wind tunnel \citep{moreau2022}, in the
anisotropic turbulence of a cylinder wake. The plate has a chord of $352$\,mm
and a perforated insert over its first $52$\,mm ($\ell/c = 0.148$), $1.5$\,mm
thick. Three square-pitch perforates were tested at two tunnel speeds and two
cylinder distances, and the local mean velocity at the plate, measured by
particle image velocimetry, is the flow speed the model uses. Noise reductions
were obtained by beamforming with a $64$-microphone array and integrating the
source map over a $0.3$\,m spanwise region centred on the leading edge. Table~\ref{tab:inputs} lists every input the model consumes. The
measured reductions are taken from their Fig.~13 at $300$--$1800$\,Hz, seven
frequencies for each of twelve curves, $84$ comparisons in all. Over this band
$k_e = 0.97$--$5.8$ and $\kb = 16$--$135$: the chord is acoustically
non-compact although $M \le 0.062$.
The wake turbulence is anisotropic: \citet{hales2024} measured a wall-normal
integral scale about twice the streamwise one, $\Lambda_2/\Lambda_1 = 1.8$--$2.4$,
and modelled it with a pseudo-anisotropic spectrum. The present model uses the
isotropic von K\'arm\'an spectrum with the measured $\Lambda_1$, which does not
affect a reduction here. The wall-normal scale integrates out of the upwash
spectrum at the plate, the amplitude and the streamwise shape of the spectrum
cancel between the rigid and permeable plates, and the span factor of
Eq.~\eqref{eq:spp} selects $k_3 \lesssim \pi/d \approx 21$\,m$^{-1}$, far below
$k_1 = \omega/U \ge 89$\,m$^{-1}$, where the spectrum barely varies with $k_3$.
Replacing the von K\'arm\'an spectrum by the anisotropic spectrum of
\citet{hales2024}, with their spectral exponent $11/3$ in place of the
von K\'arm\'an $17/6$ and their fitted spanwise scales, changes every predicted reduction by at most $0.01$\,dB.

\begin{table}[t]
\centering\small\setlength{\tabcolsep}{4pt}
\begin{tabular}{@{}lll@{}}
\toprule
quantity & value & provenance \\
\midrule
chord, insert length, thickness $h$ & $352$, $52$, $1.5$\,mm & stated \\
hole radius $R$ (cases 20, 30, 40) & $0.5$, $1.0$, $0.5$\,mm & stated \\
pitch $P$ & $3.0$, $4.0$, $1.5$\,mm & stated \\
open-area ratio $\alpha_H$ & $0.087$, $0.196$, $0.349$ & $\pi R^2/P^2$ \\
interaction function $\psi$ & $0.543$, $0.340$, $0.169$ & Eq.~\eqref{eq:fok} \\
local mean flow $U$, ($a$)--($d$) & $14.8$, $15.1$, $20.9$, $21.2$\,m\,s$^{-1}$ & measured (PIV) \\
integral scale $\Lambda_1$, ($a$)--($d$) & $14.1$, $15.2$, $13.2$, $14.5$\,mm & measured \\
grazing velocity $U_c(x)/U$ & $0.35$--$0.99$ & Eq.~\eqref{eq:uc} \\
span of the observable, $2d$ & $0.30$\,m & beamforming region \\
angular window of the array & $90^\circ\pm20^\circ$ & assumed \\
\bottomrule
\end{tabular}
\caption{Inputs for the comparison with \citet{hales2024}. No input is
adjusted to the acoustic data.}
\label{tab:inputs}
\end{table}

\subsubsection{Comparison}
\label{sec:halescmp}

\begin{figure}[t]
\centering
\includegraphics[width=\textwidth]{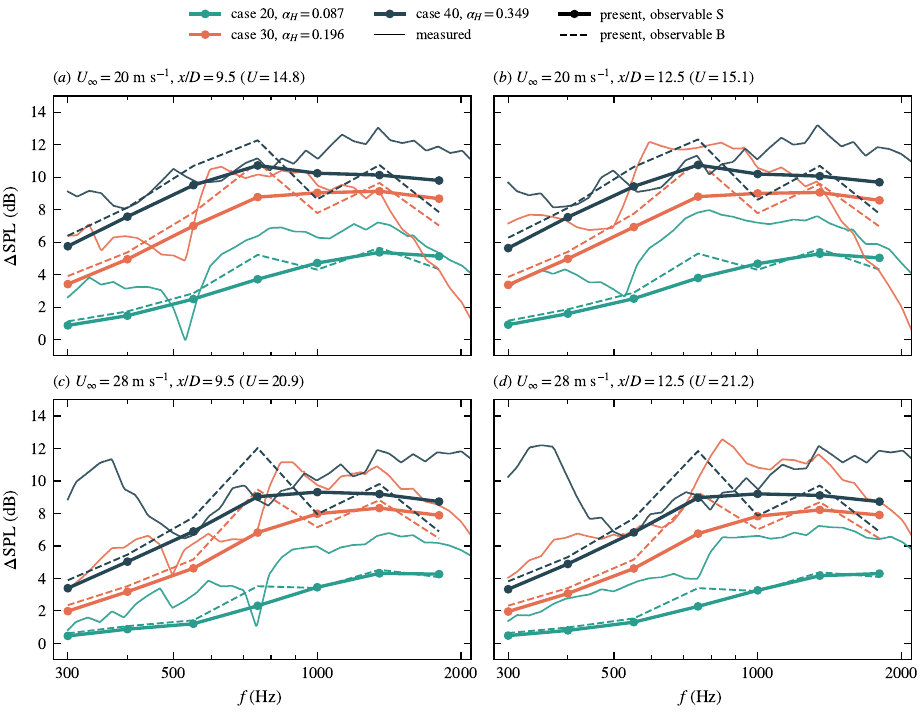}
\caption{Noise reduction of the perforated inserts of \citet{hales2024} at the
four flow conditions ($a$)--($d$). Thin lines, measurement; symbols and thick
lines, the present prediction with the span-and-array observable S; dashed,
the single-gust broadside observable B. Colours distinguish the perforates:
case~20 ($\alpha_H = 0.087$), case~30 ($0.196$) and case~40 ($0.349$). Nothing
is fitted.}
\label{fig:hales}
\end{figure}

Figure~\ref{fig:hales} compares the predicted and measured reductions for all
four flow conditions and all three perforates, and Table~\ref{tab:hales}
gives the statistics. With the span-and-array observable S the prediction agrees with the $84$
measurements to $2.46$\,dB rms with a bias of $-1.74$\,dB, an under-prediction that is nearly uniform across the perforates. The predicted ordering of the three perforates, case~20 below case~30 below
case~40, holds at every point and matches the measured ordering at $21$ of the
$28$ (condition, frequency) pairs, although $\alpha_H$ and $\psi$
enter Eqs.~\eqref{eq:KR} and \eqref{eq:Zperf} with opposite effects and the
ordering is not automatic: case~30 has half the open area of case~40 but twice
the hole radius, a larger $U_c$ and a smaller $\Dc$. All but one of the predicted curves rise from $300$\,Hz to a broad maximum
between $0.75$ and $1.4$\,kHz (the exception, case~20 at condition~$d$, rises
to the top of the band); the measured curves do the same and then fall more
steeply above $1.35$\,kHz, most strongly for case~30.
Raising the speed from condition ($a$) to ($c$), or from ($b$) to ($d$), moves
the measured rise to higher frequency, and the prediction follows because the
blockage is set by $\Sr = \omega R/U_c$ and $U_c$ scales with the local flow.

\begin{table}[t]
\centering\small
\begin{tabular}{@{}lccc@{}}
\toprule
observable & rms (dB) & bias (dB) & shape rms (dB) \\
\midrule
B: $k_3 = 0$, $\theta = 90^\circ$            & $2.45$ & $-1.51$ & $1.82$ \\
W: $k_3 = 0$, $90^\circ\pm20^\circ$            & $2.27$ & $-1.49$ & $1.58$ \\
span $0.30$\,m, $\theta = 90^\circ$            & $2.49$ & $-1.81$ & $1.58$ \\
\textbf{S: span $0.30$\,m, $90^\circ\pm20^\circ$} & $\mathbf{2.46}$ & $\mathbf{-1.74}$ & $\mathbf{1.60}$ \\
span $0.455$\,m, $90^\circ\pm20^\circ$         & $2.38$ & $-1.66$ & $1.58$ \\
\midrule
\multicolumn{4}{@{}l}{observable S by flow condition and by perforate}\\
($a$) $U = 14.8$\,m\,s$^{-1}$ & $1.91$ & $-1.01$ & \\
($b$) $U = 15.1$\,m\,s$^{-1}$ & $2.53$ & $-1.77$ & \\
($c$) $U = 20.9$\,m\,s$^{-1}$ & $2.27$ & $-1.69$ & \\
($d$) $U = 21.2$\,m\,s$^{-1}$ & $2.99$ & $-2.52$ & \\
case 20 & $2.09$ & $-1.76$ & \\
case 30 & $2.46$ & $-1.47$ & \\
case 40 & $2.79$ & $-2.00$ & \\
\bottomrule
\end{tabular}
\caption{Prediction minus measurement over the $84$ comparisons with
\citet{hales2024}, for the observables of Section~\ref{sec:observables}. The shape
rms is taken after removing the mean error of each of the twelve curves.
Compressible solution at $1.5$ times the automatic mode count.}
\label{tab:hales}
\end{table}

The choice of observable affects the spectral shape more than the overall
level. The single-gust broadside observable B has a spurious minimum near
$1$\,kHz (dashed curves in Fig.~\ref{fig:hales}), where the measurement has
none: from $750$ to $1000$\,Hz it predicts a change in reduction that is, on
average over the twelve curves, $3.2$\,dB more negative than the measured
change. The minimum arises because, at a single angle and for a single gust,
the radiation from the insert and from the rigid aft chord interfere
(Section~\ref{sec:compress}). Averaging over the array aperture reduces the error in that change to
$1.1$\,dB, and adding the finite span, through which oblique and partly
evanescent gusts contribute with different retarded phases, reduces it to
$0.3$\,dB. The shape error falls from $1.82$ to $1.60$\,dB. The overall rms
barely changes, from $2.45$ to $2.46$\,dB, because the oblique gusts carry a
smaller reduction than the normal one and lower the prediction by $0.3$--$0.7$\,dB up to $750$\,Hz, which increases the
bias from $-1.51$ to $-1.74$\,dB. The span itself is a minor uncertainty: taking the full
$0.455$\,m test-section span in place of the $0.3$\,m integration region moves
the overall statistics by less than $0.1$\,dB.

\subsubsection{Structure of the residual}
\label{sec:residual}

The residual has a systematic component. The bias is negative for all four
flow conditions and all three perforates, and it varies between the perforates
by only $0.5$\,dB ($-1.47$ to $-2.00$\,dB) although $\alpha_H$ varies
by a factor of four and $\psi$ by a factor of three. A calibration error in the
admittance would scale with the material parameters, and this one does not. By frequency the deficit is largest at the bottom of the band, $-2.9$\,dB at
$300$\,Hz, falls to between $-1.0$ and $-1.6$\,dB over $550$--$1000$\,Hz, and
is $-2.3$ and $-0.7$\,dB at $1350$ and $1800$\,Hz. The largest under-predictions, $-4.4$ to $-6.9$\,dB, are at $300$ and $400$\,Hz for case~40 at the two higher
speeds, where the measured reduction has a low-frequency hump, peaking at
$11$--$12$\,dB near $350$\,Hz, that no part of the model contains; without these four points
the remaining $80$ give $2.19$\,dB rms and $-1.56$\,dB bias.

The homogenised admittance of Section~\ref{sec:material} replaces a discrete array by
a continuum, which requires several apertures per hydrodynamic wavelength
$U/f$. At $1800$\,Hz and $U = 14.8$\,m\,s$^{-1}$ there are $2.1$ for case~30
and $2.7$ for case~20, against $5.5$ for case~40. Restricting the comparison to
$U/(fP) \ge 4$ leaves $68$ of the $84$ points and gives $2.52$\,dB rms and $-2.01$\,dB bias: the agreement is not sustained by the coarse perforates at
high frequency, nor spoiled by them. At the top of the band the measured
reductions fall, steeply for case~30. There the measurement is total noise, and the self-noise of the perforated surface, which an inviscid
scattering model cannot contain, overtakes the reduced leading-edge component;
the comparison is not extended into that range.

\subsubsection{Compressibility at a Mach number of 0.05}
\label{sec:compress}

\begin{figure}[t]
\centering
\includegraphics[width=\textwidth]{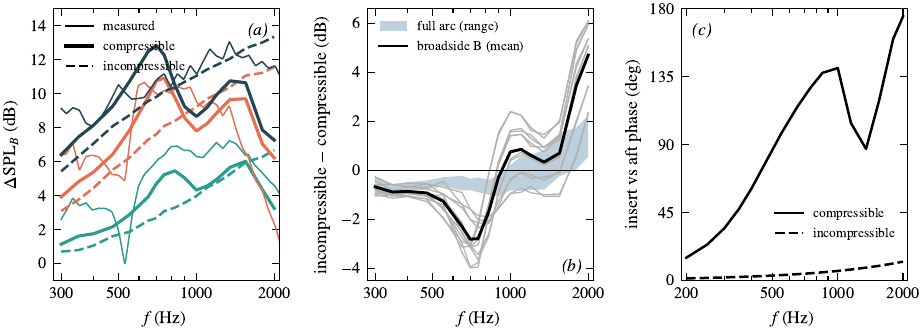}
\caption{Compressibility at $M \approx 0.05$. ($a$) Condition ($a$) of
\citet{hales2024}: measurement (thin), compressible (solid) and incompressible
(dashed) predictions with observable B. ($b$) Incompressible minus compressible
reduction for all twelve curves: broadside observable B (grey, mean in black)
and the power over the full arc (blue band). ($c$) The rigid plate at
$U = 14.8$\,m\,s$^{-1}$: phase of the insert contribution to the broadside
field relative to the aft-chord contribution, compressible and
incompressible.}
\label{fig:compress}
\end{figure}

To isolate the effect of compressibility, the same comparison was repeated with
the incompressible limit of the same solver, at the same $\kb$ and the same
$\lambda(\xi)$ and with the same radiation integral of Eq.~\eqref{eq:L}, so that the
two calculations differ only in the kernel of Eq.~\eqref{eq:sie}. Both are shown
for condition ($a$) in Fig.~\ref{fig:compress}($a$), with the single-gust
observables for which the incompressible solution is defined. Over the $84$
comparisons, observable B gives $2.45$\,dB rms and $-1.51$\,dB bias for the
compressible solution and $2.79$\,dB and $-1.66$\,dB for the incompressible one,
both at $1.5$ times the automatic mode count. The overall statistics hide a
larger difference in shape. Over $300$--$700$\,Hz the incompressible solution
under-predicts by $2.6$\,dB on average, against $1.2$\,dB for the compressible
one, and over $1.4$--$2$\,kHz it over-predicts by $1.4$\,dB where the compressible solution under-predicts by $1.5$\,dB. Between $1350$ and
$1800$\,Hz the measured reduction falls in all twelve curves, by $1.9$\,dB on
average; the compressible prediction also falls in all twelve, by $2.1$\,dB,
and the incompressible one rises in all twelve, by $1.1$\,dB. The difference
between the two solutions reaches $-4.0$ to $+5.3$\,dB on the comparison
frequencies and $+6.1$\,dB at $2$\,kHz (Fig.~\ref{fig:compress}($b$)).

The mechanism is the acoustic non-compactness of the chord. At broadside the
radiation phase in Eq.~\eqref{eq:L} is only $k_eM\xi$, so $L(90^\circ)$ is
essentially the chordwise integral of the loading, and it can be split into the
part carried by the insert, $\xi < s_j$, and the part carried by the rigid aft
chord. For the rigid plate the two parts are nearly in phase in the incompressible
solution, within $12^\circ$ up to $2$\,kHz, so the reduction is a smooth,
rising function of frequency. In the compressible solution the loading is
coupled along the chord with retarded phase $k_e|\xi - s|$, and the phase
difference grows from $35^\circ$ at $300$\,Hz to between $87^\circ$ and
$175^\circ$ above $550$\,Hz (Fig.~\ref{fig:compress}($c$)); the rigid
broadside level itself is then
a partly cancelling sum, which is why it falls $5$--$12$\,dB below the compact
value in Fig.~\ref{fig:verify}($b$), and the broadside reduction is a ratio of
two such sums. The interference that produces the spurious minimum of
observable B is the same effect seen at one angle. Integrated over the full arc,
the difference between the incompressible and compressible reductions shrinks to
between $-1.1$ and $+2.2$\,dB ($0.8$\,dB rms) over $300$--$2000$\,Hz, because the
interference moves with angle; the broadside observable exaggerates it. Compressibility therefore
changes the predicted reduction by amounts comparable to the residual of the
comparison at $M = 0.05$, and it does so through $k_e = M\kb$ rather than through
$M$ \citep[see also][]{lee2026}.

One caution applies to the top of the band. The measured roll-off moves with
flow speed: the frequency at which each measured curve has fallen $3$\,dB below
its maximum rises by a factor of about $1.2$--$1.3$ between $U \approx 15$ and
$21$\,m\,s$^{-1}$, whereas the turnover of the compressible prediction is fixed
by $k_0c$ and does not move with $U$. The agreement in the sign of the slope
above $1.35$\,kHz is therefore partly coincidental; the measured fall probably
includes the onset of self-noise, which is strongest for case~30, the perforate
with the largest holes.

\subsubsection{Isolation of the material law}
\label{sec:isolation}

Holding the solver, the observable and the data fixed and changing only the
transfer impedance isolates the material law. Table~\ref{tab:isolation} lists
the variants with observables B and W. The span-and-array observable, which
needs sixteen solutions per frequency rather than one, is evaluated for the present model only; for it, B and W give overall statistics within about $0.25$\,dB of those
of S (Table~\ref{tab:hales}).

\begin{table}[t]
\centering\small\setlength{\tabcolsep}{4pt}
\begin{tabular}{@{}lcccc@{}}
\toprule
 & \multicolumn{2}{c}{B} & \multicolumn{2}{c}{W} \\
transfer impedance & rms & bias & rms & bias \\
\midrule
\textbf{present}: $\psi$, $U_c$ from Eq.~\eqref{eq:uc} & $\mathbf{2.45}$ & $\mathbf{-1.51}$ & $\mathbf{2.27}$ & $\mathbf{-1.49}$ \\
$\psi = 1$, $U_c = U$ \citep[unscaled]{hales2024} & $4.36$ & $-3.91$ & $4.41$ & $-3.94$ \\
$\psi = 1$, $U_c$ from Eq.~\eqref{eq:uc} & $3.23$ & $-2.64$ & $3.16$ & $-2.63$ \\
$\psi$, $U_c = 0.7U$ & $2.62$ & $-1.61$ & $2.53$ & $-1.63$ \\
$\psi = 1$, $U_c = 0.7U$ & $3.39$ & $-2.83$ & $3.40$ & $-2.85$ \\
$\psi$, turbulent $U_c$ of Eq.~\eqref{eq:ucturb} & $3.02$ & $-2.31$ & $2.97$ & $-2.34$ \\
$\psi$, no grazing flow ($\Hgr = 1$) & $4.95$ & $+0.48$ & $4.82$ & $+0.85$ \\
blockage on the end corrections only & $5.03$ & $+3.56$ & $4.93$ & $+3.73$ \\
liner two-parameter model & $3.99$ & $+2.09$ & $3.80$ & $+2.24$ \\
two-parameter resistance, present reactance & $5.01$ & $+3.77$ & $4.89$ & $+3.92$ \\
\bottomrule
\end{tabular}
\caption{Isolation of the material law: prediction minus measurement (dB) over
the $84$ comparisons with \citet{hales2024}, with only the transfer impedance
changed. Compressible solution at $1.5$ times the automatic mode count.}
\label{tab:isolation}
\end{table}

The impedance of \citet{hales2024} before their empirical scaling is
Eq.~\eqref{eq:Zperf} with $\psi = 1$ and $U_c = U$, so it differs from the present
one in exactly the two computed terms. With it the prediction falls
$3.9$\,dB below the measurement on average ($4.36$\,dB rms), which is the
over-prediction of radiated noise that led \citet{hales2024} to introduce their
scaling function. Restoring the two terms removes $61\,\%$ of
that deficit with no free constant. The two contribute comparably. Replacing $U_c = U$ by the aperture
average at $\psi = 1$ moves the bias from $-3.91$ to $-2.64$\,dB, and adding
Fok's function moves it to $-1.51$\,dB, so the two terms account for
$1.3$ and $1.1$\,dB of the change.

The comparison with the conventional $U_c = 0.7U$ is instructive because its
overall statistics are close to those of the computed average
($2.62$ against $2.45$\,dB rms): over these inserts the laminar average
happens to straddle $0.7$. The difference is in how the error is distributed.
With $U_c = 0.7U$ the bias ranges from $-0.61$\,dB for case~30 to
$-2.16$\,dB for case~40; with the aperture average it ranges from $-1.21$ to
$-1.78$\,dB. The
constant cannot represent the dependence of $U_c$ on the hole radius through
$\eta_R$ in Eq.~\eqref{eq:uc}: case~30 has twice the hole radius of the other two,
and it is the case the constant treats differently. The
turbulent closure of Eq.~\eqref{eq:ucturb} gives a larger grazing velocity, a weaker
blockage at a given frequency and a larger error ($3.02$\,dB rms, $-2.31$\,dB bias) than either, consistent with the laminar state established independently
in Section~\ref{sec:bl}. Removing the grazing flow altogether gives a small mean
error for the wrong reason: the bias is $+4.0$\,dB for case~20 and $-3.8$\,dB
for case~40, because without the shed-vorticity resistance the admittance is
purely reactive and scales with $\alpha_H$ in a way the measurements do not.

The last three rows test closures from the acoustic-liner literature. Applying
Howe's blockage to the end corrections only, with the channel inertia in series,
over-predicts the reduction by $3.6$\,dB on average, which supports applying the
blockage to the whole conductivity. The two-parameter liner model
\citep{motsinger1991,yu2008}, in its linear form without a backing cavity and
with the local laminar displacement thickness, over-predicts by $2.1$\,dB on
average and by about $5$\,dB below $600$\,Hz; using only its grazing-flow
resistance with the present reactance is worse, over-predicting by
$6.4$\,dB there. These closures are
discussed in Section~\ref{sec:discussion}.

\subsection{Bulk porous inserts: Bowen et al.}
\label{sec:bowen}

\subsubsection{The experiment and the inputs}

\citet{bowen2022} measured a NACA~0012 aerofoil of $200$\,mm chord and
$600$\,mm span at $U = 20$\,m\,s$^{-1}$ ($M = 0.058$) in grid turbulence of
$10.1\,\%$ intensity and $10.8$\,mm integral length scale. The first $10\,\%$ of
the chord was replaced by an additively manufactured Schwarz-P lattice of
porosity $\phi = 40$, $50$ or $60\,\%$, whose Darcy permeability was measured
on separate samples as $\kappa = 2.78$, $3.78$ and $4.98\times10^{-9}$\,m$^2$.
The reduction is reported as the difference in power spectral density at a
single microphone directly above the leading edge at $1.75$\,m, so the
observable is B. Interaction noise dominates the measured spectra only between
about $160$ and $1000$\,Hz; outside that band facility noise and the roughness
noise of the porous surface take over.

The model uses the bulk law of Eq.~\eqref{eq:Zbulk} with the measured permeability.
The aerofoil is represented by a flat plate, its thickness entering only
through the seepage path, taken as the local thickness of the NACA~0012
section bounded below by the leading-edge diameter of $6.3$\,mm. Two inputs are
not reported and are treated as uncertainties: the tortuosity, with a baseline
$\alpha_\infty = 1.2$ and a range $1.0$--$1.6$, and the floor of the seepage
path. Halving the floor to the leading-edge radius changes the rms error in the core
band by at most $0.5$\,dB, and the diameter is retained. The reactive transition
frequency of Section~\ref{sec:bulk}, $\phi r/(2\pi\rho_0\alpha_\infty)$, is
$240$--$280$\,Hz for the three samples, so over most of the band the layer is
inertive: at $600$\,Hz, $\lambda$ at the nose is $0.22 + 0.46\I$ for
$\phi = 40\,\%$ and $0.29 + 0.73\I$ for $60\,\%$.

\subsubsection{Comparison}

\begin{figure}[t]
\centering
\includegraphics[width=\textwidth]{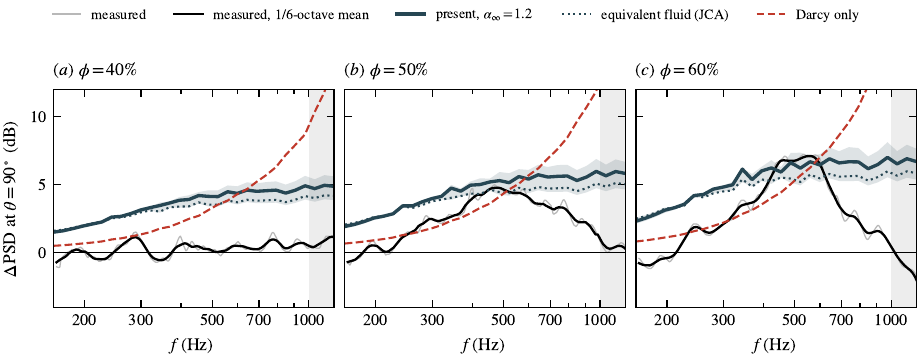}
\caption{Noise reduction of the bulk porous leading edges of \citet{bowen2022}
at $\theta = 90^\circ$. Grey, measurement; black, its one-sixth-octave running
mean; blue, the present prediction with the measured permeability and
$\alpha_\infty = 1.2$, with the band spanning $\alpha_\infty = 1.0$--$1.6$;
red dashed, Darcy resistance without the
pore-fluid inertia; dotted, the full equivalent-fluid model with the viscous characteristic length
obtained from the permeability. The shaded strip
above $1$\,kHz is outside the band where interaction noise dominates.}
\label{fig:bowen}
\end{figure}

Figure~\ref{fig:bowen} compares the predictions with the measurements, and
Table~\ref{tab:bowen} gives the statistics against a one-sixth-octave running
mean of the measured spectra. For $\phi = 50$ and $60\,\%$ the prediction agrees
with the measurement to $1.4$ and $1.7$\,dB rms over $300$--$700$\,Hz, the band
in which interaction noise is clearly dominant, and the peak reductions are
within $0.5$\,dB ($5.3$ against $4.8$\,dB, and $6.7$ against $7.1$\,dB). The
ranking with porosity is correct, and the predicted reduction grows gently with
angle from $35^\circ$ to $150^\circ$, as \citet{bowen2022} report at
$600$\,Hz. The pore-fluid inertia matters above the core band. With Darcy's resistance
alone the prediction between $300$ and $700$\,Hz is comparable, and for the
$40$ and $50\,\%$ samples slightly better, but the reduction then keeps
rising, to $9$--$16$\,dB at $1$\,kHz, far above the measurement. Over the
whole interaction-noise band, $160$--$1000$\,Hz, including the inertia lowers
the rms error from $3.4$ to $2.1$\,dB for $\phi = 50\,\%$ and from $5.0$ to
$3.0$\,dB for $60\,\%$ (Table~\ref{tab:bowen}). The tortuosity is therefore the most influential unreported input:
$\alpha_\infty = 1.6$ reduces the core-band error to $0.8$ and $1.5$\,dB rms.
Replacing Eq.~\eqref{eq:Zbulk} by the full equivalent-fluid model
\citep{johnson1987,champoux1991}, with the exact transfer impedance of a layer,
changes the core-band bias by at most $0.8$\,dB, and by less when the
viscous characteristic length is taken as the measured pore radius.

\begin{table}[t]
\centering\small\setlength{\tabcolsep}{3.5pt}
\begin{tabular}{@{}lccccc@{}}
\toprule
 & measured peak & predicted & \multicolumn{2}{c}{$300$--$700$\,Hz} & $160$--$1000$\,Hz \\
sample & (dB at Hz) & there & $\alpha_\infty = 1.2$ & $\alpha_\infty = 1.6$ & $\alpha_\infty = 1.2$ \\
\midrule
$\phi = 40\,\%$ & $1.1$ at $287$ & $3.0$ & $3.8$ / $+3.8$ & $3.2$ / $+3.2$ & $3.4$ / $+3.3$ \\
$\phi = 50\,\%$ & $4.8$ at $476$ & $5.3$ & $1.4$ / $+1.3$ & $0.8$ / $+0.5$ & $2.1$ / $+1.9$ \\
$\phi = 60\,\%$ & $7.1$ at $557$ & $6.7$ & $1.7$ / $+1.0$ & $1.5$ / $+0.0$ & $3.0$ / $+2.4$ \\
\bottomrule
\end{tabular}
\caption{Prediction minus measurement (dB), rms / bias, for the leading edges
of \citet{bowen2022}, against the one-sixth-octave running mean of the
measured spectra.}
\label{tab:bowen}
\end{table}

Two departures are systematic. Below about $250$\,Hz the prediction exceeds
the measurement by $1.6$--$3.5$\,dB for all three samples, where the measured
reduction is close to zero. Above about $700$\,Hz the measured reductions fall
while the predicted ones do not; \citet{bowen2022} attribute the
high-frequency noise increase of the porous edges to roughness noise, whose
onset this is likely to be, since surface roughness alters the boundary layer
and the edge noise it produces \citep{gill2022ice}; the model contains only the
leading-edge component.

The $\phi = 40\,\%$ sample is over-predicted by $3.8$\,dB throughout the core
band, and the measured reduction is close to zero at all frequencies. Its
measured permeability is only $26\,\%$ lower than that of the $50\,\%$ sample,
yet the measured reduction collapses. Its minimum pore diameter, however, is
$2.6$ times smaller ($0.58$ against $1.48$\,mm). The full equivalent-fluid
model, which resolves the viscous layer in the pores, raises the impedance of
this sample by only $11$--$21\,\%$ and removes $0.4$--$0.5$\,dB of the
over-prediction. To reproduce the measured reduction the impedance would have
to be four to eight times the equivalent-fluid value (the predicted reduction
at $600$\,Hz falls from $3.8$\,dB to $1.1$ and $0.5$\,dB at four and eight
times). A grazing-flow entrance impedance at the surface does not single out
this sample either, since at small aperture Strouhal number it scales with the
open fraction and not with the pore size. The most likely explanation is that
the as-built nose of the $40\,\%$ sample is much less open than the separately
tested permeability sample, for example through partly blocked throats at the
skin of the finest lattice. A homogenised model cannot settle this, and a
permeability measurement on the manufactured leading edge itself would.

\subsection{A limit of the model: perforated plates of Chaitanya et al.}
\label{sec:chaitanya}

\subsubsection{The measurements and the inputs}

\citet{chaitanya2020jsv} measured flat plates $2$\,mm thick perforated with
holes of $3$\,mm diameter at an open-area ratio of $0.32$, in grid turbulence,
and reported the reduction in sound power integrated over sixteen microphones
between $40^\circ$ and $130^\circ$ from the downstream axis at $1.2$\,m. Two
configurations are used here: plates of $150$\,mm chord at
$U = 60$\,m\,s$^{-1}$ ($M = 0.175$) perforated over the leading
$\ell/c = 0.20$, $0.32$ and $0.49$ of the chord, and fully perforated plates of
$100$ and $150$\,mm chord at $40$ and $60$\,m\,s$^{-1}$. The observable is P over
the same arc, at $k_3 = 0$ since the microphones lie in the mid-span plane of a
$450$\,mm span. The pitch of the holes is not reported; a square pitch
consistent with the open area gives $P = 4.70$\,mm, $d/P = 0.64$ and
$\psi = 0.195$ (Fig.~\ref{fig:material}($a$)). The material law is otherwise
exactly that of Section~\ref{sec:hales}. Because the holes are larger than the local
boundary layer over most of the plate, the laminar and turbulent closures give
nearly the same grazing velocity here, $U_c/U = 0.82$--$0.93$ on average over
the perforated region, so the state of the layer does not matter.

\subsubsection{Comparison}

\begin{figure}[t]
\centering
\includegraphics[width=\textwidth]{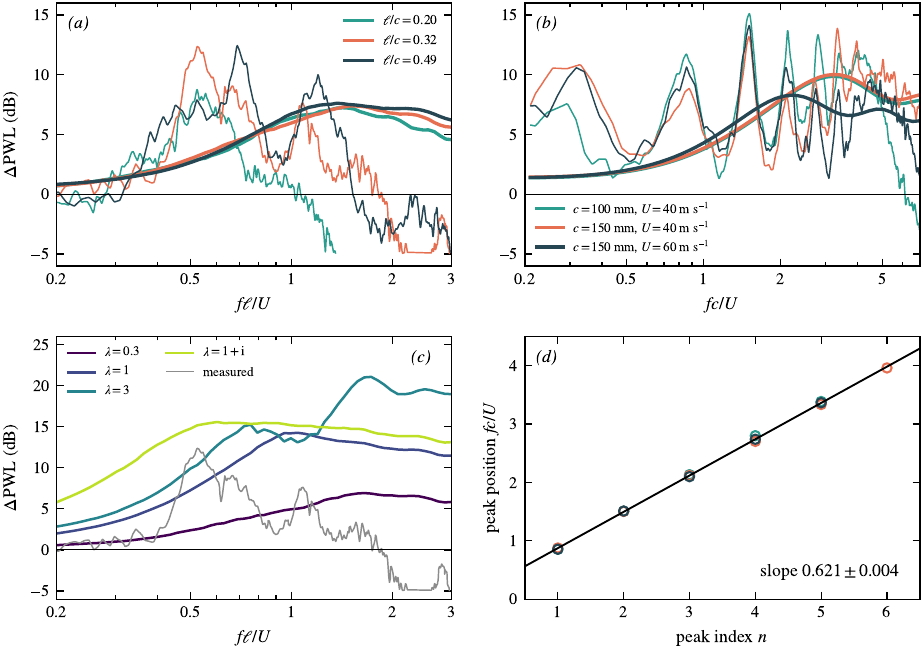}
\caption{The perforated plates of \citet{chaitanya2020jsv}. ($a$) Partial
inserts, $c = 150$\,mm, $U = 60$\,m\,s$^{-1}$: measured (thin) and predicted
(thick) power reduction against $f\ell/U$. ($b$) Fully perforated plates:
measured (thin) and predicted (thick) against $fc/U$. ($c$) The prediction for
the $\ell/c = 0.32$ insert with a constant $\lambda$ over the insert, for four
values spanning resistive and mixed surfaces. ($d$) Positions of the
measured peaks of ($b$) against their index, for the three plates; the line has
slope $0.621$.}
\label{fig:chaitanya}
\end{figure}

Figure~\ref{fig:chaitanya}($a$,$b$) compares the prediction with the
measurements. The broadband level is reproduced: over $0.25 \le f\ell/U \le
1.5$ the mean error for the three partial inserts is $+1.4$, $-0.6$ and
$-1.2$\,dB, and over the band of the peaks, $0.56 \le fc/U \le 4.2$, the mean
error for the fully perforated plates is between $-1.7$ and $-0.2$\,dB. The
spectral shape is not. The measured reduction of the partial inserts is
concentrated in narrow peaks of $8.8$, $12.4$ and $12.4$\,dB at
$f\ell/U = 0.52$--$0.69$, and falls below zero above $f\ell/U \approx 1.5$;
the prediction is a single broad maximum of $7.2$--$7.6$\,dB at
$f\ell/U = 1.4$--$1.5$. The measured reduction of the fully perforated plates
is a sequence of narrow peaks of up to $15$\,dB separated by minima close to
zero, and the prediction is again one broad maximum. The rms errors,
$3.2$--$4.7$\,dB, are therefore set by the shape. The negative reductions at the
highest frequencies are self-noise of the perforated surface, as in
Section~\ref{sec:hales}.

\subsubsection{Diagnosis}

Two tests show that the missing peaks are not a deficiency of the material
law. First, no admittance produces them. Figure~\ref{fig:chaitanya}($c$) shows
the prediction for the $\ell/c = 0.32$ insert with a constant $\lambda$ over
the insert, from weakly permeable ($\lambda = 0.3$) to strongly permeable
($\lambda = 1$ and $3$) and mixed ($\lambda = 1 + \I$).
Every curve is smooth on the scale of the measured peaks. The junction
singularity of Section~\ref{sec:sie} is sharp in space and therefore broad in
wavenumber, and since the far field samples the loading only at chordwise
wavenumbers $|\chi| \le k_e$, it radiates weakly; the scattering solution contains no
second compact source at the end of the insert that could interfere with the
leading edge at a sharply defined frequency. Measurements with the perforation
moved downstream of a rigid leading edge point the same way
\citep{palleja2022}: at a fixed open-area ratio the peak reductions were almost
unchanged when the hole diameter was varied from $1$ to $4$\,mm, and the peaks
fell at the same frequencies as those of two rigid plates separated by an open
gap, which carries no pressure jump at all.

Second, the peak positions identify what is missing.
Figure~\ref{fig:chaitanya}($d$) plots the positions of the prominent peaks of
the three fully perforated plates against their index. They fall on one line,
$fc/U = 0.25 + 0.621\,n$, to within $0.06$, for both chords and both speeds.
Equally spaced peaks in $fc/U$ are the signature of interference between the
leading edge and a disturbance that crosses the chord in the time $c/U_d$, and
the slope gives $U_d/U = 0.621 \pm 0.004$, independent of speed and chord.
\citet{chaitanya2020jsv} normalised their frequencies with a convection
velocity of $0.7U$; the data themselves indicate $0.62U$. In
the present model the only convected disturbance is the frozen gust, which
travels at $U$, and the permeable surface enters only through a local
admittance. What the slower disturbance is has not been established.
\citet{palleja2022} found peaks of the same kind on flat plates with a rigid
leading edge followed by a band of holes, at $f\ell_d/U_d \approx n$ in their
normalisation ($U_d \approx 0.7U$), where $\ell_d$ is the distance from the
leading edge to the end of the band, and, for a single row of holes at a
distance $\ell_0$ from the leading edge, at $f\ell_0/U_d \approx n$. There the
disturbance crosses rigid plate as well as perforated plate. They attributed the
peaks to secondary vorticity of opposite sign, generated at the leading edge and
convected downstream, which inverts the phase of the source at the end of the
band. A hydrodynamic wave of the boundary layer over the perforate is another
candidate. Either is a convected source that the model does not contain and that
no change of the material law can supply. Adding such a source to the
finite-chord solution is the natural extension, and it is left for future work.
Until then, the model predicts the broadband level of a perforated plate but not
the narrow-band interference that dominates its measured spectrum.

\section{Parameter study and discussion}
\label{sec:param}

The validated model is used here to isolate the parameters that control the
reduction of a perforated leading-edge insert. The baseline is the case-40
insert of \citet{hales2024} on the same plate: $c = 352$\,mm, $\ell/c = 0.148$,
$h = 1.5$\,mm, $R = 0.5$\,mm, $\alpha_H = 0.349$, at $U = 20$\,m\,s$^{-1}$ with a
laminar layer. One parameter is varied at a time (Table~\ref{tab:param}), and
each configuration is computed with the compressible solver at ten
one-third-octave frequencies from $250$ to $2000$\,Hz. The observable is the
power reduction P over the full arc, which avoids the single-angle interference
of Section~\ref{sec:compress}.

\begin{table}[t]
\centering\small\setlength{\tabcolsep}{4pt}
\begin{tabular}{@{}lll@{}}
\toprule
parameter & values (baseline in bold) & held fixed \\
\midrule
open-area ratio $\alpha_H$ & $0.02$, $0.05$, $0.10$, $0.20$, $\mathbf{0.349}$, $0.50$ & $R$; with, without $\psi$ \\
hole radius $R$ (mm) & $0.125$, $0.25$, $\mathbf{0.5}$, $1.0$, $2.0$ & $\alpha_H$ ($P = 3R$) \\
thickness $h$ (mm) & $0.5$, $1.0$, $\mathbf{1.5}$, $3.0$, $5.0$ & \\
insert length $\ell/c$ & $0.05$, $0.10$, $\mathbf{0.148}$, $0.25$, $0.50$ & \\
flow speed $U$ (m\,s$^{-1}$) & $15$, $\mathbf{20}$, $30$, $45$, $60$ & \\
\bottomrule
\end{tabular}
\caption{Configurations of the parameter study. The baseline is the case-40
insert of \citet{hales2024}; each row varies one parameter from it.}
\label{tab:param}
\end{table}

\begin{figure}[t]
\centering
\includegraphics[width=\textwidth]{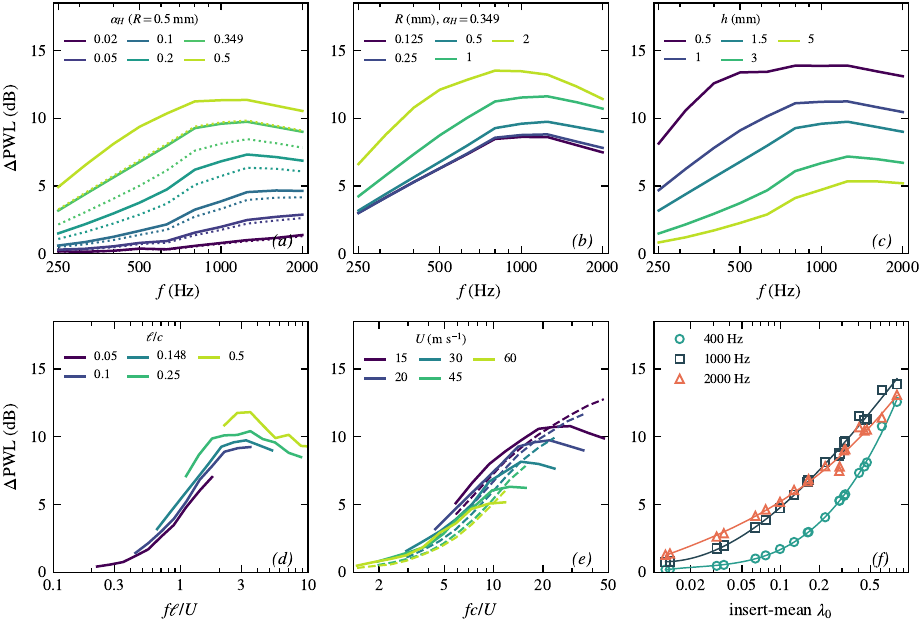}
\caption{Parameter study about the case-40 insert of \citet{hales2024}. Power
reduction over the full arc for ($a$) the open-area ratio at fixed hole radius,
with (solid) and without (dotted) Fok's function; ($b$) the hole radius at fixed
open area; ($c$) the plate thickness; ($d$) the insert length, against
$f\ell/U$; ($e$) the flow speed, against $fc/U$, compressible (solid) and
incompressible (dashed). ($f$) The reductions of the twenty configurations of
($a$)--($c$) at three frequencies against the insert-averaged low-frequency
group $\lambda_0$ of Eq.~\eqref{eq:lam0}; lines are cubic fits in $\log\lambda_0$.}
\label{fig:param}
\end{figure}

\subsection{Geometry of the perforate}

Figure~\ref{fig:param}($a$)--($c$) shows the reduction for the three geometric
parameters of the perforate. The open-area ratio dominates: at $1$\,kHz the
reduction grows from $0.8$\,dB at $\alpha_H = 0.02$ to $11.3$\,dB at $0.5$.
Fok's function adds $0.2$\,dB at $\alpha_H = 0.05$, where the holes are far
apart, and $1.1$--$1.6$\,dB at $\alpha_H = 0.2$--$0.5$, where the insert is
close-packed and $\psi$ is a significant part of $\Dc$. The plate thickness is
nearly as influential: at the same open area, thinning the plate from $5$ to
$0.5$\,mm raises the reduction at $1$\,kHz from $4.7$ to $13.9$\,dB, because the
channel inertia $2h/(\pi R)$ dominates $\Dc$. The hole radius at fixed open
area behaves differently. Below $R = 0.25$\,mm it has no effect
($8.6$--$8.8$\,dB at $1$\,kHz), and above it the reduction rises to $11.5$ and
$13.5$\,dB at $R = 1$ and $2$\,mm.

All three observations follow from the low-frequency group of Eq.~\eqref{eq:lam0}.
Figure~\ref{fig:param}($f$) plots the reductions of the twenty configurations
of panels ($a$)--($c$) against the insert average of $\lambda_0$. At each
frequency they collapse onto a single curve, with a scatter of $0.12$, $0.28$
and $0.36$\,dB rms about a smooth fit at $400$, $1000$ and $2000$\,Hz, although
$\alpha_H$, $R$, $h$ and $\psi$ enter $\lambda_0$ in different combinations.
The collapse holds because, over most of the band, the apertures operate at
$\Sr \lesssim 1$ where the admittance is close to its resistive limit
(Fig.~\ref{fig:material}($d$)). Two limits of Eq.~\eqref{eq:lam0} then explain the
dependence on hole radius. For apertures much smaller than the viscous length
$L_\nu = \sqrt{\nu x/U}$, Eq.~\eqref{eq:uc} gives $U_c/U \to F''(0)R/(2L_\nu)$, and
in a plate thick compared with the holes $\Dc \to 2h/(\pi R)$, so that
\begin{equation}
\lambda_0 \;\to\; \frac{\pi}{F''(0)}\,\frac{\alpha_H\,L_\nu}{\beta_M\,h}
         \approx 9.46\,\frac{\alpha_H\,L_\nu}{\beta_M\,h}
\qquad (R \ll L_\nu,\ h \gg R) .
\label{eq:lam0small}
\end{equation}
The hole radius cancels: a hole buried in the laminar layer sees a grazing
velocity proportional to its radius, which exactly offsets the channel
inertia $2h/(\pi R)$. For the baseline plate Eq.~\eqref{eq:lam0small} gives
$\lambda_0 = 0.29$, against $0.285$ and $0.287$ from the full expression at
$R = 0.125$ and $0.25$\,mm. For apertures much larger than the layer,
$U_c \to U$ and $\lambda_0 \to \alpha_H/[\beta_M(\psi + 2h/\pi R)]$, which
grows with $R$. The transition between the two regimes is set by $R/L_\nu$, so
the same perforate is resistive-buried near the nose of a fast, thin layer and
not further aft: the ratio of hole radius to the local boundary-layer scale is a
design parameter in its own right, and one that a constant $U_c/U$ cannot
represent.

\subsection{Insert length and flow speed}

The insert length enters differently (Fig.~\ref{fig:param}($d$)). Plotted
against $f\ell/U$, the reductions of all five inserts rise together from
$f\ell/U \approx 0.3$ and saturate beyond $f\ell/U \approx 2$, where the
insert spans two hydrodynamic wavelengths; at a given $f\ell/U$ the longer
inserts give a larger reduction, in the order of their $\lambda_0$, which grows
with $\ell$ because $U_c/U$ falls along the insert. At a fixed frequency a
longer insert therefore helps at low frequency, where a short one is
hydrodynamically compact, while above about $1.5$\,kHz lengths beyond
$\ell/c \approx 0.1$ bring nothing: at $2$\,kHz the $\ell/c = 0.1$ insert
gives $9.2$\,dB and the $\ell/c = 0.5$ insert $8.5$\,dB.

The flow speed does not collapse on the Strouhal number
(Fig.~\ref{fig:param}($e$)). At $fc/U = 10$ the reduction falls from
$8.2$\,dB at $U = 15$\,m\,s$^{-1}$ to $5.1$\,dB at $60$\,m\,s$^{-1}$. Two
effects could contribute. The boundary layer thins with speed, so $U_c/U$ rises
and $\lambda_0$ falls, from $0.34$ at $15$\,m\,s$^{-1}$ to $0.23$ at
$60$\,m\,s$^{-1}$; and at fixed $fc/U$ the acoustic wavenumber $k_e = M\kb$
grows in proportion to $M$. The incompressible solution separates them. It
shows most of the same fall, from $7.3$ to $5.1$\,dB at $fc/U = 10$, so the
speed dependence at fixed Strouhal number is mainly a boundary-layer effect,
carried by $U_c/U$ in Eq.~\eqref{eq:lam0}. Compressibility adds a correction of
either sign: the compressible reduction exceeds the incompressible one by up to
$1.2$\,dB in mid-band and falls below it at the top of the band, by $0.8$\,dB
at $60$\,m\,s$^{-1}$ and $2.9$\,dB at $15$\,m\,s$^{-1}$ at $2$\,kHz, where the
chord spans two acoustic wavelengths. A perforated insert designed at one speed therefore loses
effectiveness at a higher speed at the same Strouhal number, and the loss can be estimated from $\lambda_0$ before any scattering calculation
is made. On a rotor blade, where the local speed grows with radius, the same
argument favours a perforate that varies along the span, and the finite-chord
transfer function can be evaluated blade element by blade element, as analytic
edge-noise models already are in rotor broadband predictions
\citep{li2024serrated}.

\subsection{What the residual indicates}
\label{sec:discussion}

The comparisons of Sections~\ref{sec:hales}--\ref{sec:bowen} share a pattern. Where
the model applies, the prediction agrees with the measurement to about
$2.5$\,dB rms with no adjustment, and the remaining error has a structure. For
the perforates it is an under-prediction of about $1.7$\,dB, nearly uniform
across perforates, that is largest at the lowest aperture Strouhal numbers and
does not scale with the open-area ratio. The material law contains one term whose derivation does not match the
geometry: Howe's blockage factor of Eq.~\eqref{eq:howe} is the result for an isolated
aperture spanned by a free shear layer that can support a Kelvin--Helmholtz
wave over the aperture length. The perforates tested are $9$--$35\,\%$ open,
with webs comparable to the hole diameter, and the shear layers of neighbouring
holes are not independent. An aperture-array conductivity in grazing flow,
the grazing-flow counterpart of Fok's no-flow interaction function, does not
appear to exist in the literature. Its derivation would act on every perforate
prediction at once and is the most direct route to removing the residual.

The acoustic-liner closures of Section~\ref{sec:isolation} fail in the opposite
direction, and the reason is informative. The grazing-flow resistance of the
two-parameter model is three to eight times smaller along the insert, at $300$\,Hz, than Howe's low-Strouhal resistance, and the predicted reduction is
then too large by about $5$\,dB on average below $600$\,Hz. That model
was calibrated on facesheets under developed turbulent boundary layers at Mach
numbers of $0.3$--$0.5$, where the displacement thickness is large compared with
the hole; its dependence on $\delta^*/d$ does not carry over to a thin laminar
layer with $R \approx \delta_{99}$. The liner forms do better than the present
closure above $1$\,kHz ($1.6$--$1.9$\,dB rms against $2.65$\,dB between $1$ and $1.8$\,kHz), which suggests
that the missing array effect depends on Strouhal number in a way that none of
the closures captures without spoiling the low-frequency end. Taken with the
comparison of the laminar and turbulent grazing velocities, the conclusion is
that the grazing-flow term at a leading-edge insert must be evaluated from the
local boundary layer, and that importing a liner correlation does not supply it.

\subsection{Envelope of validity}

The material laws of Section~\ref{sec:material} are homogenised, and three limits
follow. The perforate law requires several apertures per hydrodynamic
wavelength; at $U/(fP) < 4$ the continuum assumption is marginal
(Section~\ref{sec:residual}). The bulk law is linear, which requires the pore
Reynolds number $v_sd_{\rm pore}/\nu$ to be of order unity or below; beyond it
Forchheimer's correction \citep{forchheimer1901} makes the resistance depend on
the gust amplitude. And the grazing-velocity closure must be chosen from the
flow: the laminar and turbulent averages differ by up to a factor of $1.5$ at a
leading-edge insert, and the transition criterion should be evaluated for each
configuration.

The formulation is written at zero incidence. At an angle of attack $\alpha$ the
insert carries a mean pressure difference and a mean seepage, which has no
counterpart on a rigid aerofoil. The mean seepage cannot exceed the value that
would remove the mean loading, so the mean velocity in an aperture obeys
$U_b \lesssim U\alpha/\alpha_H$. For the inserts of \citet{hales2024} at
$\alpha = 5^\circ$ and $U = 14.8$\,m\,s$^{-1}$ this bound is $3.7$--$14.8$\,m\,s$^{-1}$, comparable to or
larger than the grazing velocities of Section~\ref{sec:bl}. The bias flow is then a
competing term in the aperture Strouhal number rather than a small correction,
and the present results apply for $\alpha \ll (U_c/U)\,\alpha_H$, that is,
for angles of attack in radians well below about half the open-area ratio for
small holes and three quarters of it for large ones. Because the formulation accepts an arbitrary $\lambda(x)$, a
bias-flow conductivity \citep{howe1979,luong2005,jing2000} evaluated from a
steady porous solution can be introduced without changing the solver.

Finally, the model predicts the leading-edge component alone. In every data set
considered here the measured reduction eventually falls, and becomes negative, at
high frequency, where the self-noise of the permeable surface, generated by the
boundary layer passing over the holes or pores, overtakes the reduced
interaction noise. The frequency at which this happens moves with flow speed
and with the size of the holes, and it bounds the useful band of any
permeable treatment; predicting it requires a model of the surface-generated
noise, for example through the wall-pressure spectrum beneath the layer
\citep{lee2018wps}, which is outside the present scope.

\section{Conclusions}
\label{sec:conclusions}

The noise reduction of perforated and porous leading-edge inserts has been
predicted with a compressible finite-chord scattering solution, closed to the
far field, and material laws that take their inputs from measured geometry,
measured permeability and the local boundary layer. No constant in the
calculation was adjusted to acoustic data, and the finite span of the aerofoil
and the aperture of the measuring array were represented where the experiment
defines them.

For a perforated insert the admittance was derived from the Rayleigh
conductivity of one aperture with three computed corrections: the plate
thickness, the interaction of neighbouring apertures through Fok's function,
and Howe's grazing-flow blockage evaluated at the velocity averaged over the
aperture opening from the local boundary-layer profile. At low aperture
Strouhal number this admittance does not vanish but becomes resistive, with a
frequency-independent value set by the open area, the plate thickness, the
interaction function and the boundary layer. Against the measurements of
\citet{hales2024}, three perforates at four flow conditions, the prediction
agrees to $2.5$\,dB rms over $84$ comparisons, orders the perforates as the measurement does at $21$ of $28$ points, follows the shift of the reduction with
speed, and under-predicts by about $1.7$\,dB, nearly uniformly across
perforates and without scaling with the open-area ratio. The two computed terms,
the interaction function and the aperture-averaged grazing velocity, remove $61\,\%$ of the deficit left by the impedance used previously, and they
distribute the remaining error evenly across perforates of different hole
radius, which a constant grazing velocity does not. Closures imported from
acoustic-liner practice, calibrated on thick turbulent layers over small holes,
over-predict the reduction at a leading-edge insert.

Although the Mach number of that experiment is only $0.05$, its chord spans up
to two acoustic wavelengths, and the compressible and incompressible solutions of
the same problem differ by up to $6$\,dB at broadside. The incompressible
solution under-predicts the low-frequency reduction further than the
compressible one and rises where the measurement falls. At a single angle and a
single gust the non-compact chord also produces a spurious minimum in the
predicted reduction, which disappears when the finite span and the array
aperture are included; the choice of observable therefore matters more for
the spectral shape than for the level.

For bulk porous inserts, a Darcy resistance with pore-fluid inertia and the
measured permeability predicts the reductions measured by \citet{bowen2022} for
the $50$ and $60\,\%$ porosity samples to $1.4$ and $1.7$\,dB rms in the band
where interaction noise clearly dominates, and the inertia keeps the predicted
reduction bounded at higher frequency. The $40\,\%$ sample is over-predicted by
nearly $4$\,dB, and neither the full equivalent-fluid model nor a
grazing-flow entrance impedance explains the difference; a permeability
measured on the manufactured leading edge is needed.

The model does not reproduce the narrow, equally spaced peaks in the reduction
measured by \citet{chaitanya2020jsv} for perforated plates. None of the uniform admittances tested, resistive or mixed, produces them, and their positions
fall on a line in $fc/U$ whose slope implies a disturbance convecting over the
plate at $0.62$ of the free-stream speed, independently of speed and chord. A
frozen gust convecting at the free-stream speed and scattered by a permeable
edge cannot generate that interference. A convected source is the missing
element. Because similar peaks appear when the disturbance crosses a rigid
section ahead of the perforation \citep{palleja2022}, its origin, vorticity
generated at the leading edge or a wave of the boundary layer over the
perforate, remains to be established, and incorporating it is the natural next
step. For the
perforates, the residual under-prediction points to one missing derivation, the
conductivity of a dense aperture array in grazing flow.

Finally, the parameter study shows that for a perforated leading-edge insert the
geometry enters the reduction almost entirely through the single low-frequency
group $\lambda_0$: the reductions of twenty perforates of different open area,
hole radius, thickness and interaction collapse on it to within $0.4$\,dB rms.
For holes buried in the laminar layer the hole radius drops out, and
$\lambda_0$ is proportional to the open-area ratio times the ratio of the viscous
length to the plate thickness; larger holes help only once they protrude through
the layer. Lengthening the insert increases the reduction only until it spans
about two hydrodynamic wavelengths. And at a fixed Strouhal number the reduction
falls with flow speed, mainly because a thinner boundary layer raises the
grazing velocity and lowers $\lambda_0$; compressibility changes the reduction
by up to $3$\,dB at a given frequency even at these low Mach numbers.

\section*{CRediT authorship contribution statement}
\textbf{Seongkyu Lee:} Conceptualization, Methodology, Software, Validation,
Formal analysis, Investigation, Writing -- original draft, Writing -- review
\& editing, Visualization.

\section*{Declaration of competing interest}
The author declares that he has no known competing financial interests or
personal relationships that could have appeared to influence the work reported
in this paper.

\section*{Data availability}
The digitised measurements, the solver and the scripts that produce every
figure and table will be made available on request.

\section*{Funding}
This research did not receive any specific grant from funding agencies in the
public, commercial, or not-for-profit sectors.

\section*{Declaration of generative AI and AI-assisted technologies in the
writing process}
During the preparation of this work the author used Claude (Anthropic) to
assist with checking derivations, writing and cross-checking the numerical
scripts, and language editing. After using this tool, the author reviewed and
edited the content as needed and takes full responsibility for the content of
the publication.

\bibliographystyle{elsarticle-num-names}
\bibliography{refs}

\end{document}